\documentclass[11pt]{article}

\pdfoutput=1
\usepackage{amsmath,amssymb,epsfig,color,bbold,cite}
\usepackage{graphicx,caption,subcaption}
\renewcommand{\arraystretch}{1.2}

\usepackage{bm}
\usepackage{latexsym}
\usepackage{amssymb}
\usepackage{epsfig,amsmath,graphicx}
\usepackage{listings}

\usepackage{color}
\definecolor{Mahogany}{rgb}{0.62,0.24,0.15}
\definecolor{colorLink}{rgb}{0.7,0,0}
\definecolor{colorCite}{rgb}{0,.7,0}
\definecolor{colorURL}{rgb}{0,0,0.7}
\usepackage[colorlinks=true,linktocpage=true,linkcolor=black,citecolor=colorCite
,urlcolor=colorURL]{hyperref}

\usepackage{color}
\definecolor{myblue}{rgb}{.9, .97, .97}

\usepackage{amsmath}
\usepackage{empheq}
\usepackage{framed}

\newlength\mytemplen
\newsavebox\mytempbox

\makeatletter
\newcommand\mybluebox{%
    \@ifnextchar[
       {\@mybluebox}%
       {\@mybluebox[0pt]}}

\def\@mybluebox[#1]{%
    \@ifnextchar[
       {\@@mybluebox[#1]}%
       {\@@mybluebox[#1][0pt]}}

\def\@@mybluebox[#1][#2]#3{
    \sbox\mytempbox{#3}%
    \mytemplen\ht\mytempbox
    \advance\mytemplen #1\relax
    \ht\mytempbox\mytemplen
    \mytemplen\dp\mytempbox
    \advance\mytemplen #2\relax
    \dp\mytempbox\mytemplen
    \colorbox{myblue}{\hspace{1em}\usebox{\mytempbox}\hspace{1em}}}

\usepackage{psfrag}
\usepackage{graphicx}
\usepackage{bm}
\usepackage{feynmf}

\makeatletter
\def\endfmffile{%
  \fmfcmd{\p@rcent\space the end.^^J%
          end.^^J%
          endinput;}%
  \if@fmfio
    \immediate\closeout\@outfmf
  \fi
  \ifnum\pdfshellescape=\@ne
    \immediate\write18{mf "\noexpand\mode:=localfont;input \thefmffile"}%
  \fi}
\makeatother

\usepackage{setspace}
\usepackage{xspace}

\newcommand{\be}{\begin{equation}}  
\newcommand{\ee}{\end{equation}} 
  
\newcommand{\nl}{\nonumber \\ }

\newcommand{\order}{ {\cal O} }

\def\fmfblobEllipse#1#2#3{\fmfcmd{vblobEllipse ((#1), (#2), \fmfpfx{#3});}}

\allowdisplaybreaks

\begin{document}

\begin{titlepage}

\begin{flushright}
EFI Preprint 14-32\\
FERMILAB-PUB-14-359-T \\
SLAC-PUB-16094 
\\
September 25, 2014
\end{flushright}

\vspace{0.4cm}
\begin{center}
\Large\bf 
Soft Collinear Effective Theory for Heavy WIMP Annihilation 
\end{center}

\vspace{0.2cm}
\begin{center}
{\sc   Martin Bauer$^{1,2}$, Timothy Cohen$^{3}$, Richard J. Hill$^{1}$ and 
Mikhail P. Solon$^{1,4}$
}\\
\vspace{0.4cm}
{\it $^{1}$Enrico Fermi Institute and Department of Physics \\
The University of Chicago, Chicago, Illinois, 60637, USA
\\[2mm]
$^{2}$Theoretical Physics Department \\
Fermi National Accelerator Laboratory, Batavia, IL 60510, USA
\\[2mm]
$^{3}$Department of Physics, Princeton University, Princeton, NJ, 08544, USA \\
School of Natural Sciences, Institute for Advanced Study, Princeton, NJ, 08540, 
USA \\
Institute of Theoretical Science, University of Oregon, Eugene, OR, 97403, 
USA\\
SLAC National Accelerator Laboratory, Menlo Park, CA, 94025, USA
\\[2mm]
$^{4}$Berkeley Center for Theoretical Physics, Department of Physics\\
and Theoretical Physics Group, Lawrence Berkeley National Laboratory\\
University of California, Berkeley, CA 94270, USA
}
\end{center}
\vspace{0.2cm}
\begin{abstract}
\vspace{0.1cm}
\begin{spacing}{1.1}
\noindent   In a large class of models for Weakly Interacting Massive Particles (WIMPs), the WIMP mass $M$ lies far above the weak scale $m_W$.  This work identifies universal Sudakov-type logarithms $\sim \alpha \log^2 (2\,M/m_W)$ that spoil the naive convergence of perturbation theory for annihilation processes.  An effective field theory (EFT) framework is presented, allowing the systematic resummation of these  logarithms.  Another impact of the large separation of scales is that a long-distance wavefunction distortion from electroweak boson exchange leads to observable modifications of the cross section.  Careful accounting of momentum regions in the EFT allows the rigorous disentanglement of this so-called Sommerfeld enhancement from the short-distance hard annihilation process.   The WIMP is described as a heavy-particle field, while the electroweak gauge bosons are treated as soft and collinear fields. Hard matching  coefficients are computed at renormalization scale $\mu \sim 2\,M$, then evolved down to $\mu \sim m_W$, where  electroweak symmetry breaking is incorporated and the matching onto the relevant quantum mechanical Hamiltonian is performed. The example of an $SU(2)_W$ triplet scalar dark matter candidate annihilating to line photons is used for concreteness, allowing the numerical exploration of the impact of next-to-leading order corrections and log resummation.   
For $M \simeq 3$ TeV, the resummed Sommerfeld enhanced cross section is reduced by a factor of $\sim 3$ with respect to the tree-level fixed order result.  
\end{spacing}
\end{abstract}
\vfil

\end{titlepage}

\setcounter{tocdepth}{2}
\large
\vspace{50pt}

\tableofcontents

\pagebreak

\normalsize
\begin{spacing}{1.3}
\begin{fmffile}{fmf_paper}

\setcounter{page}{3}
\section{Introduction}

Determining the particle nature of dark matter is one of the primary
goals of the particle physics community \cite{Feng:2014uja}.  One
framework that has received tremendous attention stems from the simple
assumption that the dark matter communicates with the Standard Model
via the weak interactions.  If the Universe had a simple thermal
expansion history from temperatures of $\gtrsim \text{TeV}$ until
today,\footnote{It is entirely plausible that the history of the
Universe was more complicated in such a way that the relic density of
dark matter would be impacted \cite{Moroi:1999zb, Gelmini:2006pw,
Acharya:2009zt, Moroi:2013sla, Easther:2013nga}.  This motivates
providing results for a full range of masses as opposed to restricting
to the ``thermal" value.  Additionally, it is possible that the WIMP
is a subdominant component of the dark matter.} it is  natural for a
Weakly Interacting Massive Particle (WIMP) to freeze out with the
measured dark matter abundance (for a review, see \cite{Bertone:2004pz}).  
Another attractive
feature of WIMP models is that they lead to observable
signatures in some combination of direct detection, indirect
detection, and collider experiments.

The most studied WIMPs tend to have masses in the
$\mathcal{O}(100 \text{ GeV})$ range.  Avoiding phenomenological 
constraints while yielding the measured abundance often
requires multi-state systems that include mass mixing~\cite{Cohen:2011ec, Cheung:2012qy}, 
\emph{e.g.} the ``well-tempered neutralino" of the Minimal Supersymmetric Standard
Model (MSSM)~\cite{ArkaniHamed:2006mb}. 
Another compelling class of WIMP candidates consists of 
dark matter composed of (mostly) pure gauge eigenstates of the weak interactions.  This scenario can arise from models 
that extend the Standard Model by only minimal field 
content~\cite{Cirelli:2005uq, Cirelli:2007xd, Cirelli:2008id, Cirelli:2009uv}.  
If these WIMPs are thermal relics, then a hierarchy between the weak scale $m_W$ and the mass scale of these new particles $M$ is predicted~\cite{Cirelli:2005uq, Hisano:2006nn, Hryczuk:2010zi}.  
Additionally, the MSSM can reproduce features of this minimal dark matter
paradigm when the lightest superpartner is the pure wino or the pure Higgsino.  
Similar candidates can emerge from underlying composite structure~\cite{Hur:2007uz,Kilic:2009mi,Frandsen:2009mi,Bai:2010qg}.

The multi-TeV mass regime also becomes increasingly motivated as bounds from
collider experiments become more stringent (\emph{e.g.}~for an overview in the context of supersymmetry 
searches at LHC8, see \cite{Craig:2013cxa}).  
One interpretation of these null results is that the new physics scale will emerge somewhat
higher than the weak scale.
Clearly, WIMP models with $M\gg m_W$ deserve careful study.

From a field-theoretic point
of view, this regime becomes interesting because physical processes can exhibit
generic behavior as an expansion in the small ratio of scales, in the same manner that hydrogen-like
atomic spectroscopy or heavy meson phenomenology exhibit universal
leading order behavior in $ (\alpha\, m_e) / m_{\rm nucleus}$ or $\Lambda_{\rm
QCD}/m_{\rm heavy\,\,quark}$ respectively.  This same universality also emerges for heavy WIMP processes.  

Heavy WIMPs are difficult to probe experimentally.  Searches can be performed at the LHC, but the current mass reach is only on the order of a few hundred GeV \cite{Chatrchyan:2012tea, Chatrchyan:2012me, Aad:2013oja, Aad:2013yna, Aad:2014vka, Khachatryan:2014mma}.  Recently, it has
been shown that a future collider with $\sqrt{s} \sim 100 \text{ TeV}$
could have some impact on the parameter space of these models,
although it does not appear possible to probe masses that correspond to
thermal relics \cite{Low:2014cba, Cirelli:2014dsa}.  
Direct detection prospects for heavy electroweak dark matter are also
challenging.  A nonvanishing cross section only appears at 
loop level~\cite{Hisano:2011cs, Hill:2011be, Hill:2013hoa, Hill:2014yka}. 
Additionally, the larger mass implies a smaller number density.  To make matters worse, 
a universal amplitude-level cancellation occurs in the heavy WIMP
limit~\cite{Hill:2011be}.  The resulting $\sim 10^{-47}\,{\rm cm}^2$ 
cross section remains a target for next-generation direct detection searches,
but these experiments will have to contend with the presence of the neutrino background~\cite{Billard:2013qya}.

Fortunately, indirect detection is a viable probe of multi-TeV dark matter.
In particular, photon lines that result from WIMP annihilation can be searched for using gamma ray
telescopes.  In part, this rate is observable due to a
non-perturbative Sommerfeld enhancement to the cross section when
$\alpha_2 M \gtrsim m_W$~\cite{Hisano:2002fk, Hisano:2003ec, Hisano:2004ds, Cirelli:2007xd, Ciafaloni:2010ti, Hryczuk:2011vi}, where $\alpha_2$ is the electroweak fine structure constant.  
Investigation of constraints from current experiments such as H.E.S.S.~\cite{Abramowski:2013ax} 
indicate that under certain assumptions on the galactic dark matter halo model, 
some heavy WIMPs are already severely constrained from annihilation to line 
photons~\cite{Cohen:2013ama, Fan:2013faa, Hryczuk:2014hpa}.  These conclusions depend both on the halo model
and on the precise determination of the low-velocity WIMP annihilation
cross section.   While the former remains a subject of astrophysical
study, the latter lies firmly in the domain of particle physics.  

The study of such heavy WIMP annihilation processes presents a 
multi-scale field theory
problem, involving large corrections $\sim \!\alpha_2\, \text{log}^2\, (2\,M/m_W)$ in the perturbative expansion.  
A complete scale separation is desirable
both to obtain robust numerical predictions for
the cross section and to identify the universal features of heavy WIMP
annihilation.  In particular, it will be demonstrated that the dominant effect of perturbative corrections is the reduction of the 
tree-level amplitude by a universal factor.  The dominant contribution to this universal factor can be traced to  the so-called cusp anomalous dimension~\cite{Beneke:2009rj,Becher:2003kh,Becher:2009qa,Becher:2009kw},
which governs the renormalization of Wilson loops in gauge theory. 

\begin{figure}[h!]
\begin{centering}
\includegraphics[width=.65 \textwidth]{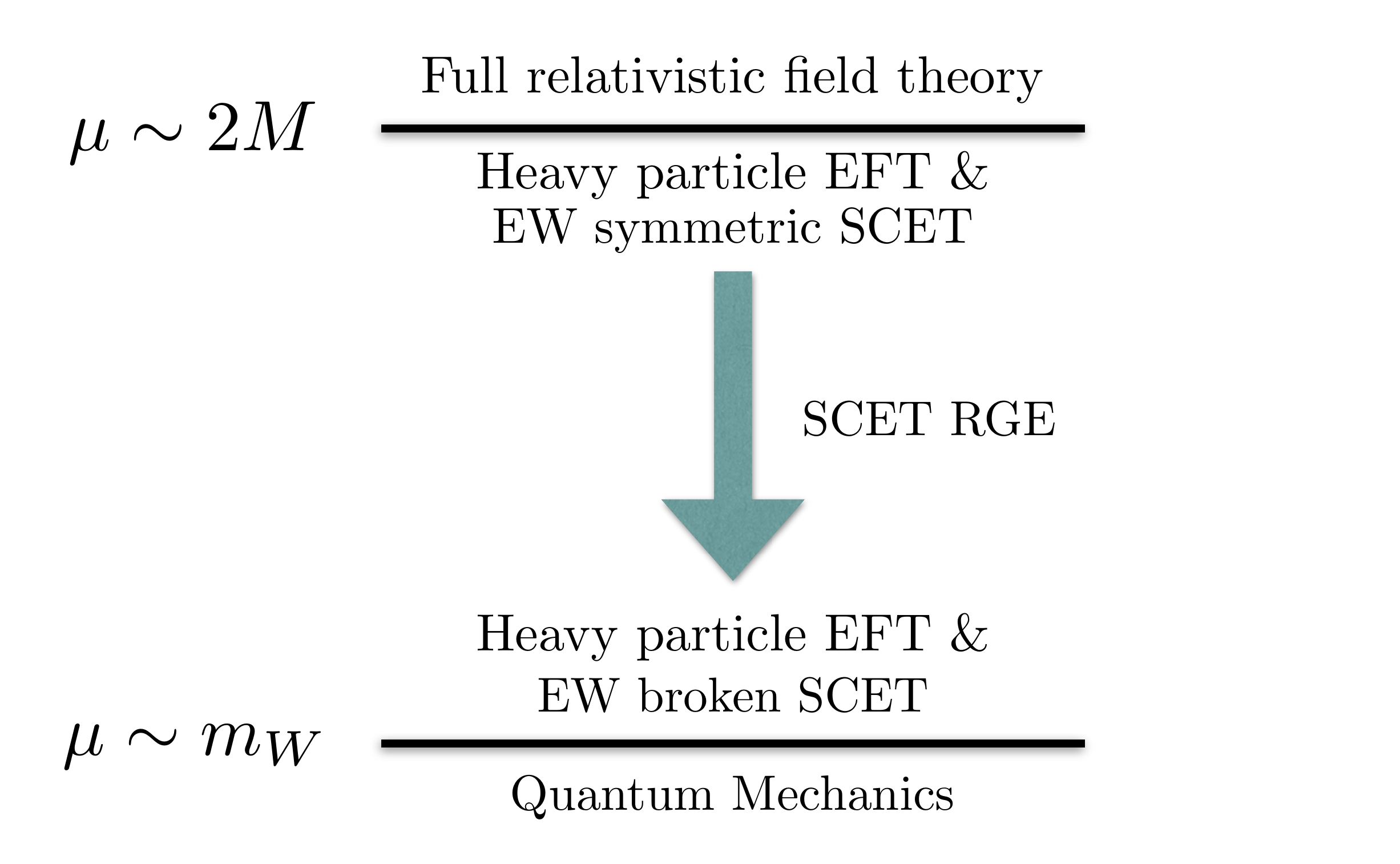}
\caption{A schematic of the EFT decomposition utilized in this calculation.}
\label{fig:scales}
\end{centering}
\end{figure}

The annihilation amplitudes can be analyzed in an Effective Field
Theory (EFT) at the operator level.
Schematically, the leading operators take the form
\be
O_\text{ann} \sim \phi_v\, \phi_v \, \mathcal{A}_n \mathcal{A}_{\bar{n}},
\ee
where $\phi_v$ and $\mathcal{A}_n$ are 
EFT fields that describe the initial state non-relativistic WIMPs and 
the final state energetic collinear electroweak gauge bosons
($v$ and $n$, $\bar{n}$ are associated timelike and lightlike vectors; 
detailed expressions are given in \eqref{eq:originalbasis} below).  
Four separate field theories are necessary to
capture the relevant physics, as sketched in Fig.~\ref{fig:scales}.  At
renormalization scales $\mu \gg M$, the full relativistic Standard
Model with the addition of the WIMP sector is appropriate.
Below $\mu \sim M$, the dynamics of the heavy WIMPs
is captured by matching onto Heavy Particle 
EFT~\cite{Caswell:1985ui, Eichten:1989zv, Neubert:1993mb, Manohar:2000dt}.
Since momentum modes with $p^2 \gtrsim M^2$ are no longer present in
the theory, final state particles must be restricted to have virtuality
small compared to this scale.  As will be demonstrated by isolating various
regions of 1-loop diagrams, the language of Soft Collinear Effective
Theory (SCET) \cite{Bauer:2000ew, Bauer:2000yr, Bauer:2001ct,
Bauer:2001yt, Bauer:2002uv, Hill:2002vw, Chay:2002vy, Beneke:2002ph} 
captures the relevant IR dynamics of the effectively massless Standard Model fields.
Given that $\mu \gg m_W$, it is  appropriate to treat the theory in
the electroweak symmetric phase, which simplifies calculations.
Next, the Renormalization Group Equation (RGE) is solved, yielding the Wilson coefficients at the scale $\mu
\sim m_W$.  At this
scale, electroweak symmetry breaking is relevant, and the appropriate
finite corrections are computed using
SCET in a field basis with broken electroweak symmetry.  
This procedure systematically resums large logarithms, providing
a controlled perturbative expansion. 
The final matching step determines the parameters of 
a quantum mechanical Hamiltonian in which 
phenomenological observables may be straightforwardly computed.  
The EFT approach allows a rigorous definition of, and separation 
between, the long-distance physics associated with 
wavefunction distortions, \emph{i.e.}, the Sommerfeld enhancement, and the
short-distance physics 
of the annihilation process.
Subleading perturbative, power, and velocity corrections may be systematically
incorporated. 

In this paper, we focus for simplicity on the case of heavy scalar triplet WIMP
annihilation to photons.  The scalar triplet can be taken as a viable dark matter candidate
 on its own, or seen as a scalar proxy for the fermionic ``wino".  The wino appears as the lightest superpartner 
in models that involve anomaly mediated supersymmetry breaking~\cite{Randall:1998uk, Giudice:1998xp}, 
and is often the dark matter candidate in models of Split 
Supersymmetry~\cite{Wells:2004di, ArkaniHamed:2004fb, Giudice:2004tc, Pierce:2004mk, 
Arvanitaki:2012ps, ArkaniHamed:2012gw, Hall:2012zp, Kane:2011kj, Ibe:2011aa}.  The analysis can be readily extended to describe heavy WIMPs of other
spins and other electroweak quantum numbers, to describe different final states, and to 
compute thermal relic abundances in addition to present-day indirect detection signatures.  
Details involving the phenomenologically interesting case of wino annihilation to line photons will be presented in future work~\cite{future}.

The remainder of the paper is structured as follows.   In
Sec.~\ref{sec:scalarmodel}, we specify the scalar model.  In
Sec.~\ref{sec:QM}, we provide the low-energy quantum mechanical
Hamiltonian and compute matching conditions in terms of  free-particle
annihilation amplitudes through one-loop order.  This will reveal
large logarithms in the matching coefficients that will be later
resummed via Renormalization Group (RG) evolution.  In
Sec.~\ref{sec:regions}, we perform a regions  analysis of
prototypical diagrams appearing in the perturbative evaluation of
heavy WIMP annihilation, and  introduce the relevant formalism of
SCET.  Section~\ref{sec:hardmatching}
gives hard matching conditions for a heavy scalar WIMP.
Section~\ref{sec:RG} derives anomalous dimensions and renormalization
group evolution equations  governing the intermediate theory at scales
$m_W \lesssim \mu \lesssim 2M$.    Section~\ref{sec:lowmatch} computes
matching conditions onto the low-scale quantum mechanical  theory.
Section~\ref{sec:pheno} gives the results for resummed physical
annihilation cross sections including the Sommerfeld enhancement and investigates the impact of resummation.
Section~\ref{sec:summary} provides a summary and outlook. 

\section{Scalar Model\label{sec:scalarmodel}}
The goal of this paper is to 
construct and apply 
an EFT appropriate for heavy WIMP annihilation.  While the formalism is general, 
for concreteness, we will 
consider a 
scalar electroweak triplet with zero hypercharge.  Consider the Lagrangian for a heavy scalar triplet, 
\be\label{eq:symmetricL}
{\cal L} = \frac12 (D_\mu\phi)^2 - \frac12 M^2 \phi^2 \,. 
\ee
The covariant derivative is 
\be\label{eq:covderiv}
iD_\mu = i\partial_\mu + g_2 W^a_\mu T^a \,,  
\ee
where $(T^a)^{bc} = i \epsilon^{bac}$ are $SU(2)_W$ generators in the 
adjoint representation. 
In the basis of electric charge eigenstates we have
\be
i D_\mu = i\partial_\mu + e Q A_\mu + \frac{g_2}{c_W} \big(T^3 - s_W^2 Q \big) 
Z_\mu + 
\frac{g_2}{\sqrt{2}} 
\big( T^+ W^+_\mu + T^- W^-_\mu\big) \,,
\ee
where $Q\equiv T^3 + Y$ is the electric charge in units of
the proton charge.  
The Lagrangian in this basis becomes 
\begin{align}\label{eq:scalarL}
{\cal L} &= \frac12 \big(\partial_\mu \phi_0\big)^2 -\frac12 M^2 \phi_0^2 
+ \partial_\mu \phi_+ \partial^\mu \phi_- - M^2 \phi_+ \phi_- 
- i g_2 W^+_\mu \big( \phi_- \partial_\mu \phi_0 - \phi_0 \partial_\mu \phi_- 
\big) 
\nl&\quad 
- i g_2 W^-_\mu \big( -\phi_+ \partial_\mu \phi_0 + \phi_0 \partial_\mu \phi_+ 
\big) 
+ ie A_\mu \big ( \phi_- \partial^\mu \phi_+ - \phi_+ \partial^\mu \phi_- \big) 
+ i g_2 c_W Z_\mu \big( \phi_- \partial^\mu \phi_+ - \phi_+ \partial^\mu \phi_- 
\big) 
\nl&\quad 
+ \phi_+ \phi_- \big( e^2 A_\mu A^\mu + 2 e g_2 c_W A_\mu Z^\mu + g_2^2 c_W^2 
Z_\mu 
Z^\mu \big) 
- \phi_0 \big( \phi_- W_\mu^+ + \phi_+ W_\mu^- \big ) \big(eg_2 A^\mu + g_2^2c_W 
Z^\mu \big) 
\nl&\quad 
 -\frac12 g_2^2 \Big[ \big(\phi_-\big)^2 W^+_\mu W^{+\mu} + \big(\phi_+\big)^2 
W^-_\mu W^{-\mu} \Big] 
+ g_2^2 \big( \phi_0^2 + \phi_+ \phi_- \big) W^+_\mu W^{-\mu} \,,
\end{align} 
from which it is straightforward to read off the Feynman rules.
Since we will be working to leading order in the small ratio $m_W/M$ and leading loop order, 
we neglect renormalizable self-couplings of the scalar field, 
$\sim \phi^4$, 
and Higgs interactions, $\sim H^\dagger H \phi^2$.  
It would be straightforward to include these couplings in an extended analysis.

\section{Fixed Order Matching onto Quantum Mechanics\label{sec:QM}}
To begin, 
let us match the WIMP annihilation process computed 
directly in the high scale 
field theory onto a quantum mechanical Hamiltonian.  
This will make clear 
the separation between the hard annihilation process and the wavefunction distortion. 
The former arise from offshell momentum regions of loop diagrams, and are represented by 
contributions to contact interactions in the quantum mechanical Hamiltonian.  The latter emerge from nearly onshell momentum 
regions, and are reproduced by corresponding quantum mechanical potentials.

The general quantum mechanical Hamiltonian appropriate for the center-of-mass 
frame for the two-particle system takes the form%
\footnote{
See \emph{e.g.} \cite{Hill:2000qi}.  A related formalism for treating velocity corrections in WIMP annihilation 
is given in \cite{Beneke:2012tg,Hellmann:2013jxa}. 
}
\be\label{eq:H} 
H = {p^2\over 2 M_r} + \Delta +  V + i W \,,
\ee
where $V$ and $W$ are Hermitian, $M_r$ denotes reduced mass, and $\Delta$ is the 
residual mass matrix, which captures the difference in rest mass energy between the 
states of interest.  
In matrix 
notation, acting on two components in the neutral-neutral ($00$) and 
charged-charged ($+-$) 
sectors, the kinetic energy and residual mass terms are
\begin{align}
\label{eq:QMKineticTerm}
{p^2\over 2M_r} + \Delta 
&= p^2 \left(\begin{array}{cc}{1\over M_0} & 0 \\ 0 & {1\over M_\pm} 
\end{array}\right) 
+ \left(\begin{array}{cc} 0 & 0 \\ 0 & 2\delta \end{array}\right) 
\,, 
\end{align}
where the zero of energy is taken as $2M_0$ and we define $\delta=M_\pm - M_0$. 
For notational convenience we will set $M_0\equiv M$ in the following. 
The potential $V+iW$ is determined by comparing  the Born series computed  
from this Hamiltonian, 
\be
\langle \bm{k}^\prime | T | \bm{k} \rangle 
= \langle \bm{k}^\prime | V+iW | \bm{k} \rangle + \dots \,, 
\ee
with the field theory prediction for the scattering amplitude. 

\subsection{Determining $V$}

The Hermitian potential $V$ will capture the effects of the long range force experienced by the WIMPs, 
and $W$ will encode the hard annihilation process via the optical theorem as 
discussed in Sec.~\ref{sec:FullThy}.  Employing the Feynman rules for heavy 
scalars from (\ref{eq:scalarL}), the result for $V$ reads
\begin{align}\label{eq:V}
\langle \bm{k}^\prime | V | \bm{k} \rangle 
&= \left(\begin{array}{cc} 0 & -4\pi \alpha_2\left[ {1 \over (\bm{k}^\prime - 
\bm{k})^2 + 
m_W^2} 
+ {1 \over (\bm{k}^\prime + \bm{k})^2 + m_W^2} 
\right]
\\
- 4\pi \alpha_2 \left[ {1 \over (\bm{k}^\prime - \bm{k})^2 + m_W^2} 
+ {1 \over (\bm{k}^\prime + \bm{k})^2 + m_W^2} 
\right]
& \quad - 4 \pi \alpha \left[ {1\over (\bm{k}^\prime - \bm{k})^2 + m_\gamma^2} + {t_W^{-2} 
\over 
(\bm{k}^\prime-\bm{k})^2 + m_Z^2} \right ]
\end{array} \right) \,,
\end{align}
where $\alpha_2=g_2^2/4\pi$ and $\alpha=e^2/4\pi$ are the electroweak and electromagnetic fine structure constants, 
$m_W$ and $m_Z$ are the $W^\pm$ and $Z^0$ boson masses, and 
$m_\gamma$ is an infinitesimal photon mass that is used to regulate IR divergences.    
In the quantum field theory calculation, the two terms in the off-diagonal 
elements of (\ref{eq:V}) arise from crossed and uncrossed diagrams involving $W^\pm$ exchange, and the terms in the lower right entry are from photon and 
$Z^0$ exchange, respectively.  Equation (\ref{eq:V}) will be used in the old-fashioned perturbation theory analysis, presented in Sec.~\ref{sec:WQM} below, 
in order to determine the correct matching onto quantum mechanics at one-loop order. 

\subsection{The Sommerfeld Enhancement}
In order to compute the Sommerfeld enhancement, it is useful to Fourier transform $V$ from (\ref{eq:V}) into position space, 
\begin{align}\label{eq:Vr} 
V^{S\text{-wave}} 
&= \left(\begin{array}{cc} 0 & - \sqrt{2} {\alpha_2 \over r} e^{-m_W r} 
\\
- \sqrt{2}   {\alpha_2 \over r} e^{-m_W r} 
& \quad - {\alpha \over r} - {\alpha_2 c_W^2 \over r} e^{-m_Z r} 
\end{array} \right) \,,
\end{align}
where this result is appropriate for $S$-wave scattering states (at $m_\gamma = 0$).  
Then this matrix can be used as the input to the $S$-wave Schr\"odinger
equation to model the wavefunctions of the neutral and charged WIMP
pairs, yielding the Sommerfeld enhancement.  Specifically, we use the
formalism outlined in the Appendix of \cite{ArkaniHamed:2008qn} to
compute the physical annihilation cross section from quantum
mechanics, using (\ref{eq:H}) as an input.   Indices $i,j =
1, 2$ refer to the $(00), (+-)$ states respectively.  For the 
wavefunction $(\psi^i)_j$, the index $i$
labels the asymptotic state and $j$ is the 
component index
for the resulting solution.  Given a choice of $i$, the boundary
conditions employed are
\begin{eqnarray} 
(\psi^i(0))_j &\rightarrow& \delta_{ij}, \quad j =
1,2\,,\\ (\psi^i(\infty))_1 &\rightarrow& e^{i k_i r} \,,\\ (\psi^i(\infty))_2
&\rightarrow&   \left \{\begin{array}{lr} \psi^i_\text{Coulomb} & : E
\geq \delta_i\\ e^{-k_i r} & : E < \delta_i
  \end{array}\,,\right.
\end{eqnarray}
where $k_i=M\,\sqrt{1-\delta_i/E}$, $E$ is the kinetic energy of the WIMP system, 
$\delta_i$ is only non-zero when $i = 2$, and $\psi^i_\text{Coulomb}$ is the wavefunction 
for the Coulomb scattering solution that depends on momentum $k_i$.\footnote{Note that to achieve numerical stability, 
we furthermore strip off the asymptotic, plane-wave or Coulomb, factors as outlined in the Appendix 
of \cite{Cohen:2013ama}.}  Once the solutions $\psi$ have been obtained, the Sommerfeld 
enhancement matrix is given by 
\be
s_{ij} = (\psi^i(\infty))_j.
\ee
The cross section can then be computed using 
\be
\sigma_i v = -2 \sum_{j,j'} s_{ij}\, W^{S\text{-wave}}_{j j'}\, s^*_{i j'},
\ee
where $W^{S\text{-wave}}$ denotes the absorptive part of the potential for $S$-wave scattering
states.%
\footnote{
For the contact interaction $W$, this amounts to the replacements $W_{11}\to W_{11}/2$, 
$W_{12}\to W_{12}/\sqrt{2}$, $W_{21}\to W_{21}/\sqrt{2}$, $W_{22}\to W_{22}$ starting 
from the plane wave basis (\ref{eq:Wtree}).}

The couplings and masses are defined as their onshell values. 
In particular, here we are using the shorthand 
$\alpha_2 = \alpha s_W^{-2}$ with $s_W^2=1-c_W^2$ and $c_W=m_W/m_Z$. 
All that is required to determine an annihilation cross section are  
(Particle Data Group~\cite{Agashe:2014kda}) inputs for 
$\alpha$, the $W^\pm$ and $Z^0$ masses, the WIMP mass $M$, the
charged-neutral mass splitting $\delta$, the relative velocity $v$, and the $2\times2$ Hermitian
matrix $W$.  
Now that the formalism for calculating the wavefunction factors has been explained, 
we move to the determination of the hard-annihilation contribution to the potential 
$W$ through one-loop order by matching field theory onto quantum mechanics. 

\subsection{Determining $W$: Full Theory}
\label{sec:FullThy}
The most straightforward way to determine the absorptive part of the potential, 
$W$, from field theory is through use of the optical theorem.  Matching is done 
at a convenient kinematic point, specifically the two-particle threshold for 
neutral or charged WIMPs for diagonal elements of $W$, or at the two-particle 
charged WIMP threshold for off-diagonal elements (such that the amplitude 
describes an onshell physical process). 

The discontinuity arising from two-photon 
final states is found to be
\begin{align}\label{eq:optical} 
i{\rm Disc}{\cal M_{\rm NR}}\Big([\phi\phi]_{i} \to [\phi\phi]_f\Big)
&= 
\parbox{40mm}{
      \begin{fmfgraph*}(100,60)
        \fmfleftn{l}{2}
        \fmfrightn{r}{2}
        \fmftopn{t}{3}
        \fmfbottomn{b}{3}
        \fmf{plain}{l2,v,l1}
        \fmf{plain}{r2,w,r1}
        \fmf{phantom}{v,w}
        \fmffreeze
        \fmf{photon,left}{v,w}
        \fmf{photon,right}{v,w}
        \fmf{dashes}{t2,b2}
        \fmfblob{0.15w}{v}
        \fmfblob{0.15w}{w}
      \end{fmfgraph*}
    }
\nl
&= -{1\over 8\pi} {1\over (\sqrt{2E_i})^2 (\sqrt{2E_f})^2} {\cal 
M}\Big([\phi\phi]_{i} \to \gamma\gamma\Big) {\cal 
M}\Big([\phi\phi]_{f}\to \gamma\gamma\Big)^* \,,
\end{align}
where the factors $\sqrt{2E}$ for each external particle convert to 
nonrelativistic state normalization (denoted by subscript ``NR''), 
and we have introduced the reduced amplitude, 
${\cal M}\Big([\phi\phi]_{i} \to \gamma(\epsilon) \gamma(\epsilon^\prime)\Big) = 
\epsilon^*\cdot \epsilon^{\prime *} {\cal M} \Big([\phi\phi]_{i} \to \gamma\gamma\Big)$. 
Identifying ${\rm Disc}{\cal M} = 2i{\rm Abs}{\cal M}$
gives the absorptive contribution from field theory.%
\footnote{For a single channel, the absorptive part is identified with the 
imaginary part, 
${\rm Abs}{\cal M} \equiv {\rm Im}{\cal M}$. 
}  

\begin{figure}[h!]
\begin{center} 
\input{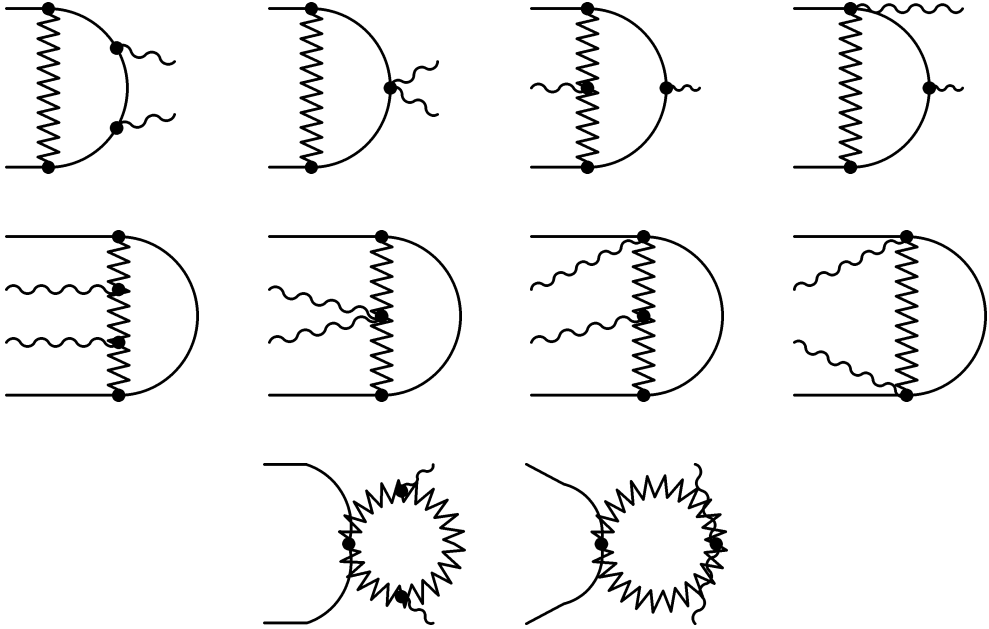}
\caption{Diagrams contributing to hard scale matching for neutral WIMPs.  Wavy 
lines are photons, zigzag lines are $W^\pm$ bosons.
}
\label{fig:diagrams_oneStep_neutral} 
\end{center}
\end{figure} 

\begin{figure}
\begin{center} 
\input{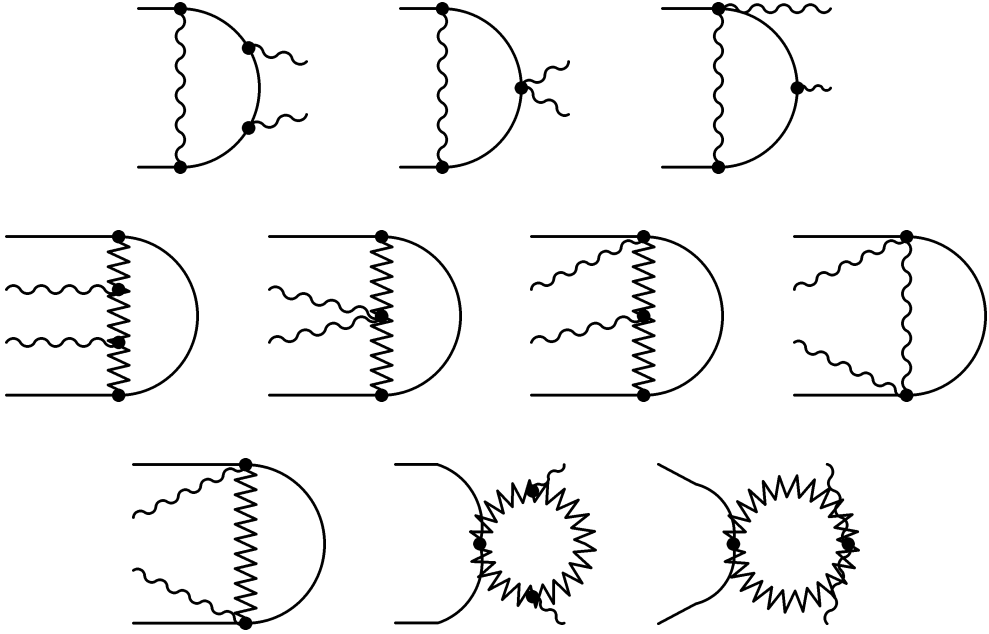}
\caption{Diagrams contributing to matching for charged WIMPs.  Wavy lines are photons, zigzag lines are $W^\pm$ bosons, and the inclusion of diagrams where internal photon lines are replaced by $Z^0$ boson lines is implied.
}
\label{fig:scalardiagrams_charged} 
\end{center}
\end{figure}

For neutral WIMP annihilation, the relevant amputated loop diagrams are shown 
in Fig.~\ref{fig:diagrams_oneStep_neutral}.  Considering kinematics at both the neutral 
and charged WIMP thresholds, we have
\begin{align} \label{eq:onestep1PI}
& {\cal M}^{00\to\gamma\gamma}
= 
{e^2 g_2^2\over (4\pi)^2}
\bigg\{ 
C_\text{potential}
+ (16 - 16 i \pi) \log{m_W\over 2M} - 4 - \pi^2 
+ {86\pi \over 3} {m_W\over M} 
\nl
&\quad
+ {m_W^2\over M^2} \bigg[ 26\log^2{m_W\over 2M} + (20+10i\pi) \log{m_W\over 2M} 
- {104\over 3} - {15\pi^2\over 2} 
-{7i\pi\over 2} \bigg] 
+ \order\left( \alpha, {\delta \over m_W}, \sqrt{\delta \over M}, {m_W^3 \over M^3} \right)
\bigg\} \,,
\end{align}
where $C_\text{potential}$ depends on whether the matrix element is evaluated at the neutral or charged WIMP threshold:
\be
\label{eq:Cpotential}
C_\text{potential} = 
\left\{
  \begin{array}{lr}
   \frac{16\pi M }{m_W + \sqrt{2M\delta}} &\quad \quad \text{for}\,\, (p+p^\prime)^2=4M_0^2\\
   {16\pi M \over \sqrt{2M\delta}}\arctan\left(\sqrt{2M\delta}\over m_W \right) 
& \quad \quad \text{for}\,\,   (p+p^\prime)^2=4M_\pm^2
  \end{array}
\right. \,.
\ee
We have here ignored higher order corrections involving the mass splitting 
(\emph{cf.}~(\ref{eq:MassSplitting}) below). 
For charged WIMP annihilation, the process has a tree-level contribution. 
Including the tree vertex with counterterms, together with the 
loop diagrams of Fig.~\ref{fig:scalardiagrams_charged}, 
\begin{align}
\label{eq:onestep1PICharged}
&{\cal M}^{+-\to\gamma\gamma}\big|_{ (p+p^\prime)^2=4M_\pm^2} = 
 Z_2^\phi (Z_1^W)^2 (Z_2^W)^{-2} 2e^2
+ { e^2 g_2^2\over (4\pi)^2} 
\bigg\{ {8\pi c_W^2 M\over m_Z} + 
{8\pi s_W^2 M\over m_\gamma} 
\nl
&\quad 
+ 8\left( c_W^2 \log{m_Z\over 2M} + s_W^2 \log{m_\gamma \over 2M} \right) 
-16 \log^2{m_W\over 2M} -16\log{m_W\over 2M} 
- 8i\pi \log{m_W\over 2M} + {3\pi^2\over 2} - 18 
\nl
&\quad 
+{m_W \over M}
\bigg[ -4\pi  + {7\pi\over 3} c_W \bigg] 
+{m_W^2\over M^2}
\bigg[ 5\log^2{m_W\over 2M} - 12 \log{m_W\over 2M} - 2\log{m_Z\over 2M}  
\nl
&\quad
+ 5i\pi \log{m_W\over 2M} 
- 12\log{2} + {20 \over 3} - {5\pi^2\over 4} - {7i\pi \over 4} 
\bigg] 
+ \order\left(\alpha, m_\gamma, {\delta \over m_W}, \sqrt{\delta \over M}, {m_W^3 \over M^3} \right)
\bigg\} \,.
\end{align}
The renormalization constant $Z_2^\phi$ is inherited from 
the electroweak symmetric Lagrangian (\ref{eq:symmetricL}) and 
$Z_1^W$, $Z_2^W$ are field and coupling renormalization factors for 
the $SU(2)_W$ gauge field~\cite{Hollik:1988ii}.%
\footnote{
Following the conventions of \cite{Hollik:1988ii}, bare Lagrangian fields and
parameters are given by $(W^a_\mu)^{\rm bare} = (Z_2^W)^{1/2} W^a_\mu$, $g_2^{\rm 
bare} = Z_1^W (Z_2^W)^{-3/2} g_2$. 
} 

Let us briefly review the renormalization for the scalar triplet. 
The 1PI two-point functions for the charged and neutral scalar fields 
at one-loop order are given by
\begin{align}
-i\Sigma_2^{\phi_+}(p^2) &= [c_\epsilon] \bigg\{ 
{2(2-\epsilon) \over \epsilon(1-\epsilon)} 
\bigg[ e^2 (m_\gamma^2)^{1-\epsilon} + g_2^2c_W^2 (m_Z^2)^{1-\epsilon} + g_2^2 
(m_W^2)^{1-\epsilon} \bigg]
\nl
&\quad 
+ {1\over \epsilon(1-\epsilon)} \bigg[ 
e^2 \left( -2 (m_\gamma^2)^{1-\epsilon} + (M^2)^{1-\epsilon} \right) 
+ g_2^2 c_W^2 \left( -2 (m_Z^2)^{1-\epsilon} + (M^2)^{1-\epsilon} \right)
\nl
&\quad 
+ g_2^2 \left( -2 (m_W^2)^{1-\epsilon} + (M^2)^{1-\epsilon} \right)
\bigg]
-{1\over \epsilon} (M^2)^{-\epsilon} \bigg[
 e^2 (2p^2 + 2M^2 -m_\gamma^2) I(m_\gamma/M,p^2/M^2) 
\nl
&\quad 
+ g_2^2 c_W^2 (2p^2 + 2M^2 - m_Z^2) I(m_Z/M,p^2/M^2) 
\nl
&\quad 
+ g_2^2 (2p^2 + 2M^2 - m_W^2) I(m_W/M,p^2/M^2) 
\bigg]
\bigg\} \,,
\nl
-i\Sigma_2^{\phi_0}(p^2) &= [c_\epsilon] g_2^2 
\bigg\{ {4(2-\epsilon)\over \epsilon (1-\epsilon)} (m_W^2)^{1-\epsilon}
+ {2\over \epsilon (1-\epsilon)} \bigg[ -2 (m_W^2)^{1-\epsilon} + 
(M^2)^{1-\epsilon} \bigg] 
\nl
&\quad 
- {2\over \epsilon} (M^2)^{-\epsilon} (2p^2 + 2M^2 - m_W^2) I(m_W/M,p^2/M^2) 
\bigg\} \,,
\end{align}
where we introduce the shorthand $[c_\epsilon] = i(4\pi)^{-2+\epsilon} \Gamma(1+\epsilon)$, and
\begin{align}
I(m,p^2) &= 
\int_0^1 dx \, [ -x(1-x)p^2 + x + (1-x)m^2 ]^{-\epsilon} 
\nl
&= 1 -\epsilon\bigg[ m\sqrt{4-m^2}\arctan\left( \sqrt{4-m^2}\over m \right) + 
m^2\log{m} - 2 \bigg]
+ \order\big(\epsilon^2\big)
\,.
\end{align}
From these results, it is straightforward to derive the one-loop 
expressions 
for the mass splitting, 
\begin{align} 
\label{eq:MassSplitting}
M^2_{\pm} - M^2_0 &= \Sigma_2^{\phi_+}\left(M_\pm^2\right) - \Sigma_2^{\phi_0}(M_0^2) 
=  \alpha\, m_W M {1 - c_W \over s_W^2 }  + \order\Big(\alpha^2, 1/M^3\Big) \,,
\end{align}
and the residue of the charged propagator
\begin{align}\label{eq:chargedres}
Z_2^{\phi_+} - 1 &= {\partial \Sigma_2^{\phi_+} \over \partial p^2 
}\bigg|_{p^2=M^2} 
= i g_2^2 [c_\epsilon] M^{-2\epsilon} \bigg[ -{4\over \epsilon} 
+ 4 \bigg( s_W^2 \log{m_\gamma \over M} + c_W^2 \log{m_Z\over M} + \log{m_W\over 
M} \bigg) 
\nl
&\quad
- {\pi m_W\over M} (1+c_W)  -  {3 m_W^2\over M^2} \bigg( \log{m_W\over M} + 
\log{m_Z\over M} \bigg)
+ \order(1/M^3, \epsilon)
\bigg] \,. 
\end{align}
Finally, for the combination of renormalization constants $(Z_1^W)^2 
(Z_2^W)^{-2}$ appearing in 
(\ref{eq:onestep1PI}), we have
\begin{align}\label{eq:ZW}
2 \delta Z_1^W - 2\delta Z_2^W 
&= -{2\over s_W c_W} {\Sigma^{AZ}(0) \over m_Z^2} 
= 
-{g_2^2 \over (4\pi)^2} m_W^{-2\epsilon} {4\over \epsilon} 
\,. 
\end{align}
In particular, $\Sigma^{AZ}(0)$ receives contributions only from the $W^\pm$ boson 
loop, and is independent of the additional scalar triplet.   

The amplitudes (\ref{eq:onestep1PI}), (\ref{eq:onestep1PICharged}) and renormalization constants 
(\ref{eq:chargedres}), (\ref{eq:ZW})
determine the physical one-loop amplitudes for heavy scalar annihilation to 
photons in terms of physical parameters $\alpha$, $m_W$, $m_Z$, $M$, $\delta$.  
One can see from these equations that there are factors of the type
$M/m_W$ that result from the so-called potential region of the loop
integrals.  It is exactly these factors that are resummed by including
the Sommerfeld enhancement.  Isolating the hard annihilation
contribution to the $W$ matrix from terms that derive from the potential
region requires working to higher order in quantum mechanics.  This is the
subject of the next section, where the equivalent quantum mechanics calculation is
performed.

\subsection{Determining $W$: Quantum Mechanics \label{sec:WQM}} 

In this section, the matching conditions for the absorptive part of the 
potential $W$ are computed in quantum mechanics. 
Working in the plane wave basis, we write
\begin{align}\label{eq:Wtree}
\Big\langle \bm{k}^\prime \Big| W^{(\gamma)} \Big| \bm{k} \Big\rangle 
&\equiv
\left(  \begin{array}{cc} 
w_{00}^{(\gamma)} & w_{00;\pm}^{(\gamma)} \\
w_{\pm;00}^{(\gamma)} & w_\pm^{(\gamma)} 
\end{array} 
\right)
\,,
\end{align}
where $w_{\pm;00}=w_{00;\pm}^*$\,, and 
the superscript $(\gamma)$ denotes restriction to $\gamma\gamma$ final states.
We work through lowest non-vanishing order in $\alpha$ for each of the elements 
$w^{(\gamma)}_{ij}$, 
but will also retain the first subleading term for $w_{\pm}^{(\gamma)}$ so that 
our computation 
contains complete one-loop corrections (see (\ref{eq:onestepw}) for explicit expressions).  
Working in the framework of ``old-fashioned" 
perturbation theory, the nonrelativistic scattering amplitude is given by the Born 
series for the matrix valued potential of (\ref{eq:H}).  What follows is the 
explicit computation of these matrix elements. In the following, we restrict to $\gamma \gamma$ final states and omit the superscript on $w_{ij}$.

For the {\bf charged channel}: 
\begin{align}
& {}_\pm\!\langle \bm{k}^\prime | T | \bm{k} \rangle_\pm  \to 
\parbox{40mm}{
      \begin{fmfgraph*}(100,60)
        \fmfleftn{l}{2}
        \fmfrightn{r}{2}
        \fmf{plain}{l2,v,l1}
        \fmf{plain}{r2,v,r1}
        \fmffreeze
        \fmfblob{0.1w}{v}
      \end{fmfgraph*}
    }
+ 
\parbox{40mm}{
      \begin{fmfgraph*}(100,60)
        \fmfcmd{input vblobEllipse}
        \fmfleftn{l}{2}
        \fmfrightn{r}{2}
        \fmf{plain,tension=2}{l2,v,l1}
        \fmf{plain}{r2,v,r1}
        \fmffreeze
        \fmfblob{0.1w}{v}
        \fmffreeze
        \fmf{phantom}{v,x,r1}
        \fmf{phantom}{v,y,r2}
        \fmffreeze
        \fmf{phantom}{x,z,y}
        \fmfblobEllipse{.4w}{.2}{z}
      \end{fmfgraph*}
    }
+ 
\parbox{40mm}{
      \begin{fmfgraph*}(100,60)
        \fmfcmd{input vblobEllipse}
        \fmfleftn{l}{2}
        \fmfrightn{r}{2}
        \fmf{plain,tension=2}{r2,v,r1}
        \fmf{plain}{l2,v,l1}
        \fmffreeze
        \fmfblob{0.1w}{v}
        \fmffreeze
        \fmf{phantom}{v,x,l1}
        \fmf{phantom}{v,y,l2}
        \fmffreeze
        \fmf{phantom}{x,z,y}
        \fmfblobEllipse{.4w}{.2}{z}
      \end{fmfgraph*}
    }
+ \dots 
\nl
&= i W_{\pm}  +   i W_\pm \otimes V_{\pm} + V_{\pm} \otimes i W_\pm + 
\order\big(\alpha^4\big)
\nl
&= 
{}_{\pm}\!\langle \bm{k}^\prime | iW | \bm{k} \rangle_{\pm}  
+ 
\int {d^3p \over (2\pi)^3} \int {d^3p^\prime \over (2\pi)^3} \,
 {}_{\pm}\!\langle \bm{k}^\prime | V | \bm{p} \rangle_{\pm} \, 
_{\pm}\langle \bm{p} | (E - H_0)^{-1} | \bm{p}^\prime \rangle_{\pm} \,
_{\pm}\langle \bm{p}^\prime | iW | \bm{k} \rangle_{\pm}  
\nl
&\quad 
+ 
\int {d^3p \over (2\pi)^3} \int {d^3p^\prime \over (2\pi)^3} \,
 {}_{\pm}\!\langle \bm{k}^\prime | iW | \bm{p} \rangle_{\pm} \, 
_{\pm}\langle \bm{p} | (E - H_0)^{-1} | \bm{p}^\prime \rangle_{\pm} \,
_{\pm}\langle \bm{p}^\prime | V | \bm{k} \rangle_{\pm}  
\nl
&= 
i w_\pm + \int {d^3 p \over (2\pi)^3} \bigg\{ 
 i w_\pm \left( {\bm{k}^2 \over M_\pm} - {\bm{p}^2\over M_\pm}\right)^{-1}
(-4\pi \alpha) \bigg[ { 1 \over (\bm{p}-\bm{k})^2 + m_\gamma^2} + { t_W^{-2}  
\over (\bm{p}+\bm{k})^2 + 
m_Z^2} \bigg] 
\nl
&\quad + 
(-4\pi \alpha) \bigg[ { 1 \over (\bm{p}-\bm{k}^\prime)^2 + m_\gamma^2} + { 
t_W^{-2}  \over (\bm{p}+\bm{k}^\prime)^2 + 
m_Z^2} \bigg] 
 \left( {\bm{k}^2 \over M_\pm} - {\bm{p}^2\over M_\pm}\right)^{-1}
i w_\pm 
\bigg\}
 \,. 
\end{align}
Here the circular blob denotes insertion of $iW$, while the elliptical blob denotes insertion of $V$. 
For neutral particle production at threshold, $\bm{k}=\bm{k}^\prime=0$, 
this gives
\begin{align}\label{eq:Tpm}
 {}_\pm\!\langle \bm{k}^\prime | T | \bm{k} \rangle_\pm 
\to i w_\pm + 2i \alpha w_\pm M_\pm \bigg( {1\over m_\gamma} + {t_W^{-2} \over 
m_Z} \bigg)  + \order(\alpha^3) \,,
\end{align}
where $m_\gamma$ is a photon mass regulating IR divergences.  

For the {\bf mixed channel}:
\begin{align}
& {}_\pm\!\langle \bm{k}^\prime | T | \bm{k} \rangle_{00}  \to 
\parbox{40mm}{
      \begin{fmfgraph*}(100,60)
        \fmfleftn{l}{2}
        \fmfrightn{r}{2}
        \fmf{plain}{l2,v,l1}
        \fmf{plain}{r2,v,r1}
        \fmffreeze
        \fmfblob{0.1w}{v}
      \end{fmfgraph*}
    }
+ 
\parbox{40mm}{
      \begin{fmfgraph*}(100,60)
        \fmfcmd{input vblobEllipse}
        \fmfleftn{l}{2}
        \fmfrightn{r}{2}
        \fmf{plain,tension=2}{r2,v,r1}
        \fmf{plain}{l2,v,l1}
        \fmffreeze
        \fmfblob{0.1w}{v}
        \fmffreeze
        \fmf{phantom}{v,x,l1}
        \fmf{phantom}{v,y,l2}
        \fmffreeze
        \fmf{phantom}{x,z,y}
        \fmfblobEllipse{.4w}{.2}{z}
      \end{fmfgraph*}
    }
+ \dots 
\nl
&= i W_{\pm;00}  + i W_\pm \otimes V_{\pm;00} + \order(\alpha^4)
\nl
&= i w_{\pm;00} 
+ 
\int {d^3 p \over (2\pi)^3} i w_{\pm} \left( {\bm{k}^2 \over M_0} - 
{\bm{p}^2\over M_\pm} - 2\delta \right)^{-1}
(-4\pi\alpha s_W^{-2})\bigg[ {1\over (\bm{p}-\bm{k})^2 + m_W^2} + {1\over 
(\bm{p}+\bm{k})^2 + 
m_W^2} \bigg] \,.
\end{align} 
Evaluated at the threshold for charged particle production, $\bm{k}^\prime=0$ and 
$\bm{k}^2 = 2M_0\delta$, this expression yields
\begin{align}\label{eq:Tmix}
 {}_\pm\!\langle \bm{k}^\prime | T | \bm{k} \rangle_{00}  &\to 
iw_{\pm;00} + 2i\alpha s_W^{-2} w_\pm {1\over \sqrt{2 M_0\delta} 
}\arctan\left(\sqrt{2M_0\delta}\over m_W\right)
+ \order(\alpha^4)\,. 
\end{align}
For the {\bf neutral channel}: 
\begin{align}
& {}_{00}\!\langle \bm{k}^\prime | T | \bm{k} \rangle_{00}  \to 
\parbox{40mm}{
      \begin{fmfgraph*}(100,60)
        \fmfleftn{l}{2}
        \fmfrightn{r}{2}
        \fmf{plain}{l2,v,l1}
        \fmf{plain}{r2,v,r1}
        \fmffreeze
        \fmfblob{0.1w}{v}
      \end{fmfgraph*}
    }
+ 
\parbox{40mm}{
      \begin{fmfgraph*}(100,60)
        \fmfcmd{input vblobEllipse}
        \fmfleftn{l}{2}
        \fmfrightn{r}{2}
        \fmf{plain,tension=2}{l2,v,l1}
        \fmf{plain}{r2,v,r1}
        \fmffreeze
        \fmfblob{0.1w}{v}
        \fmffreeze
        \fmf{phantom}{v,x,r1}
        \fmf{phantom}{v,y,r2}
        \fmffreeze
        \fmf{phantom}{x,z,y}
        \fmfblobEllipse{.4w}{.2}{z}
      \end{fmfgraph*}
    }
+ 
\parbox{40mm}{
      \begin{fmfgraph*}(100,60)
        \fmfcmd{input vblobEllipse}
        \fmfleftn{l}{2}
        \fmfrightn{r}{2}
        \fmf{plain,tension=2}{r2,v,r1}
        \fmf{plain}{l2,v,l1}
        \fmffreeze
        \fmfblob{0.1w}{v}
        \fmffreeze
        \fmf{phantom}{v,x,l1}
        \fmf{phantom}{v,y,l2}
        \fmffreeze
        \fmf{phantom}{x,z,y}
        \fmfblobEllipse{.4w}{.2}{z}
      \end{fmfgraph*}
    }
\nl
&\quad 
+
\parbox{40mm}{
      \begin{fmfgraph*}(100,60)
        \fmfcmd{input vblobEllipse}
        \fmfleftn{l}{2}
        \fmfrightn{r}{2}
        \fmf{plain}{r2,v,r1}
        \fmf{plain}{l2,v,l1}
        \fmffreeze
        \fmfblob{0.1w}{v}
        \fmffreeze
        \fmf{phantom}{v,x,l1}
        \fmf{phantom}{v,y,l2}
        \fmf{phantom}{v,a,r1}
        \fmf{phantom}{v,b,r2}
        \fmffreeze
        \fmf{phantom}{x,z,y}
        \fmf{phantom}{a,c,b}
        \fmfblobEllipse{.4w}{.2}{z}
        \fmfblobEllipse{.4w}{.2}{c}
      \end{fmfgraph*}
    }
+ \dots 
\nl
&= i W_{00}  +  V_{00;\pm} \otimes i W_{\pm;00} + i W_{00;\pm} \otimes 
V_{\pm;00} 
+ V_{00;\pm} \otimes iW_{\pm} \otimes V_{\pm;00} + \order(\alpha^5)\,.
\end{align}
Evaluating this expression at the neutral threshold, $\bm{k}=\bm{k}^\prime=0$, 
yields
\begin{align}\label{eq:T0}
{}_{00}\!\langle \bm{k}^\prime | T | \bm{k} \rangle_{00}  &\to 
iw_{00} + 4 i \alpha s_W^{-2} {M_\pm \over m_W + \sqrt{2M_\pm \delta}} 
\text{Re}\big( w_{\pm;00}  \big) 
+ \bigg[  2 \alpha s_W^{-2} {M_\pm \over m_W + \sqrt{2M_\pm \delta}} \bigg]^2 i 
w_{\pm} 
+ \order(\alpha^5) \,. 
\end{align}
Note that $T = -{\cal M}_{\rm NR}$ in the conventions employed here.%
\footnote{The source of the minus sign is simply that in the Lagrangian, $L = -V$, 
while the scattering matrix is defined as $T= + V + \dots$. 
}
The elements of $W$ are obtained by 
applying (\ref{eq:optical}), 
being careful to convert from plane-wave to $S$-wave external 
states. Equations (\ref{eq:Tpm}), (\ref{eq:Tmix}) and (\ref{eq:T0}) give the 
absorptive part of the 
non-relativistic amplitudes, which should be set equal to the corresponding 
relativistic amplitudes 
using 
the appropriate combinations of (\ref{eq:onestep1PI}) and (\ref{eq:onestep1PICharged}).
Neglecting power corrections, 
\begin{empheq}[box={\mybluebox[5pt]}]{align}
\label{eq:onestepw}
w_{\pm} &= -{\pi\alpha^2\over M^2}\bigg\{ 1 
+ {\alpha s_W^{-2}\over 4\pi} \bigg[ -16 \log^2{m_W\over 2M} - 8 \log{m_W\over 
2M} + {3\pi^2\over 2} -18
\bigg] 
\bigg\}
+ \order(\alpha^4, m_W/M) 
\,,
\nl
w_{\pm;00} &= -{\pi\alpha^2\over M^2}{\alpha s_W^{-2}\over 4\pi} 
\bigg[ (8-8i\pi) \log{m_W\over 2M} - 2 - {\pi^2\over 2} \bigg] + 
\order(\alpha^4, m_W/M) 
\,,
\nl
w_{00} &= -{\pi\alpha^2\over M^2}\left(\alpha s_W^{-2} \over 4\pi\right)^2
\bigg[ \left( 8\log{m_W\over 2M} - 2-{\pi^2\over 2} \right)^2 + 64\pi^2 
\log^2{m_W\over 2M} 
\bigg] + \order( \alpha^5, m_W/M ) \,.
\end{empheq}
Note the presence of the $\log^2( m_W/2M)$ factor (and its large coefficient)
in the one-loop correction to $w_\pm$.  
This large perturbative correction results 
in a numerically large suppression of WIMP cross 
sections compared to tree-level predictions.  
This motivates introducing an EFT that can separate the scales 
$2M$ and $m_W$ in order to resum this (and other) logarithms, thereby 
systematically improving the convergence of perturbation theory.

Power corrections in $m_W/M$ to the matching coefficients $w_{ij}$ may be obtained 
by expanding the amplitudes (\ref{eq:onestep1PI}), (\ref{eq:onestep1PICharged}).  
In the $M \gtrsim {\rm TeV}$ mass regime, these corrections 
are numerically subleading compared to logarithmically enhanced perturbative corrections at leading power~\cite{future}.

\subsection{Fixed Order Results}

\begin{figure}[top]
\centering
\includegraphics[width=.65 \textwidth]{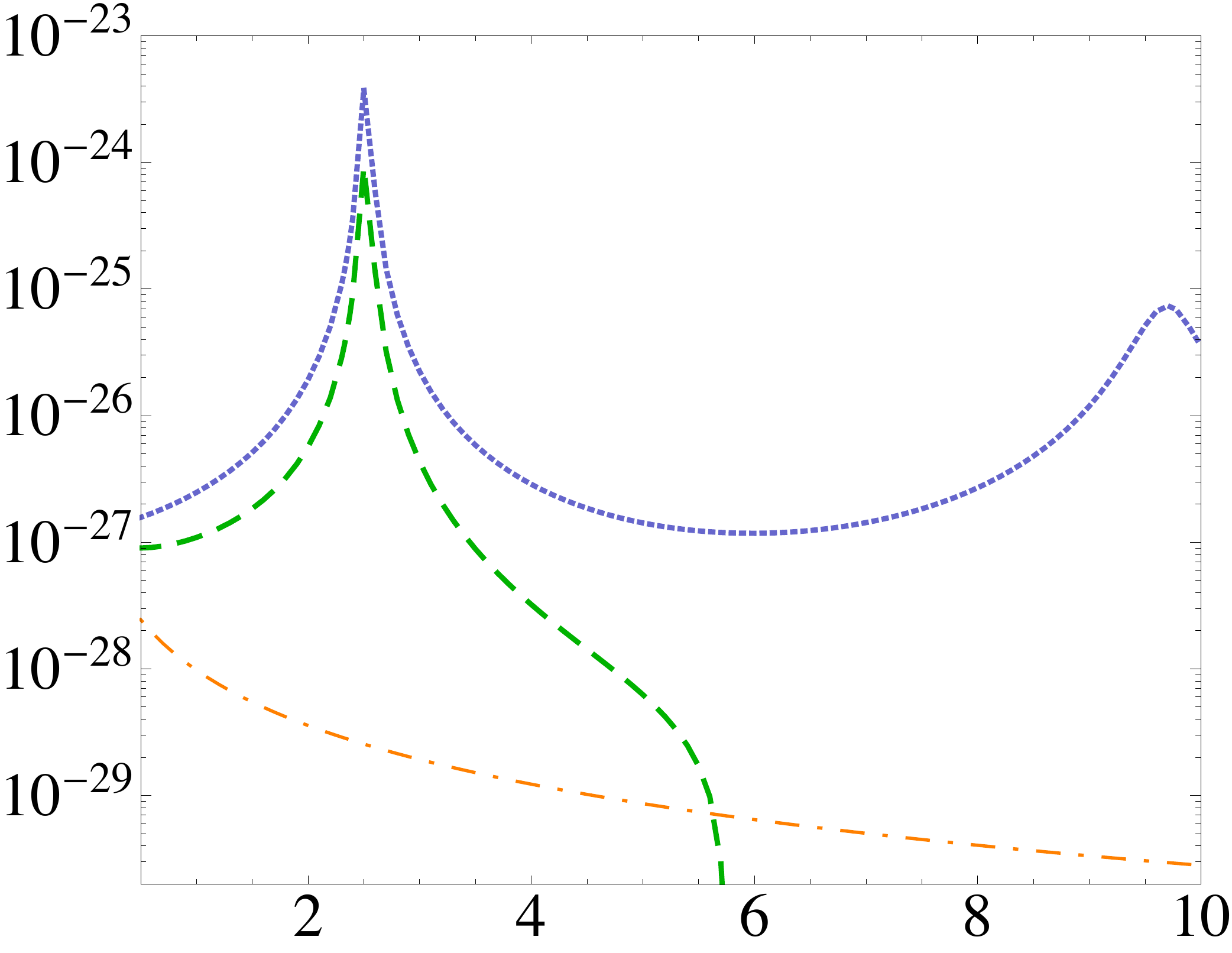}
\hspace*{10mm}
\begin{minipage}{0cm}
\vspace*{1.6cm}\hspace*{-7.1cm}{\Large $M\,\,\big[\text{TeV} \big]$}
\vspace*{0.5cm}
\end{minipage}
\begin{minipage}{0cm}
\vspace*{-8.5cm}\hspace*{-13.2cm}\rotatebox{90}{{\Large $\sigma\,v\,\,\big[\text{cm}^3/\text{s}\big]$}}
\end{minipage}
\vspace{-15pt}
\caption{Sommerfeld enhanced annihilation cross sections for two fixed order approximations. The blue dotted curve truncates the $w$ factors at $O(\alpha^2)$, while the green dashed line is the one-loop result including $O(\alpha^3)$ contributions in $w_\pm$ and $w_{\pm00}$ and the first non-vanishing $O(\alpha^4)$ contribution in $w_{00}$. Note that for $M \gtrsim 6\, \text{TeV}$, the one-loop cross section becomes negative due to the presence of a large Sudakov logarithm with a negative coefficient. For illustration we include the orange dot-dashed line which gives the naive cross section computed from $w_{00}$ neglecting wavefunction enhancements. In this plot $v = 10^{-3}$ and $\delta = 0.17~{\rm GeV}$.}
\label{fig:treeVsNLO}
\end{figure}

Armed with the Sommerfeld matrix $s_{ij}$, and the elements of the $W$
matrix given in (\ref{eq:onestepw}), we are in a position to compute
the dark matter annihilation cross section to photons at both
tree level (by simply truncating the $\alpha$ expansion in
(\ref{eq:onestepw})) and one loop.  The results of these two calculations
are shown in Fig.~\ref{fig:treeVsNLO}, where we have taken $\delta =
0.17\,\text{GeV}$ and the relative velocity $v = 10^{-3}$ in the
numerical evaluation  of the Sommerfeld enhancement.  Clearly the one-loop result is suppressed with respect to the tree-level result.  Specifically,
we find that at $M = 3\,\text{ TeV}$ (a mass of interest for the
thermal wino), the ratio $\sigma_\text{tree}/\sigma_\text{1-loop} \sim 5$.  
However the perturbative expansion is not under control, as seen 
from the fact that the fixed order $\alpha^3$ cross section becomes
negative for $M \gtrsim 6 \,\text {TeV}$ due to the large Sudakov logarithm. The orange dot-dashed line gives the naive cross section computed from $w_{00}$ neglecting the Sommerfeld enhancement.   

These considerations motivate introducing an EFT description in order
to separate the scales $m_W$ from $2M$ and resum the large logarithms, 
regaining control over the perturbative expansion.  
The first step will be
to derive an appropriate EFT description that captures all of the
relevant momentum regions of the full theory.  This is the topic of
the next section.

\section{Deriving the Effective Theory \label{sec:regions}}

In the interesting regime of large mass, the cross section becomes uncertain 
due to large Sudakov logarithms,  $\sim \alpha \log^2(m_W/2M)$.   We wish to develop an EFT framework that will isolate these
enhanced contributions and systematically reorganize the perturbative expansion 
to resum them. The framework will also reveal certain 
universal features, including properties that are independent of the WIMP's spin or 
electroweak gauge representation, and simplify matching calculations at the hard scale $\mu \sim 2 M$ and weak scale $\mu \sim m_W$; \emph{e.g.},   the hard matching can be performed using electroweak symmetric Feynman rules. 

This problem shares some
features with processes involving electroweak vector boson production
at colliders.  However, one important difference is the presence of a
heavy gauge-charged initial state in addition to jets of collinear
charged final states, in contrast to the simpler Sudakov problem
involving gauge-singlet heavy particle production \cite{Chiu:2007yn,
Chiu:2007dg, Chiu:2008vv, Manohar:2014vxa}.  The problem also shares
some features with heavy particle pair production such as $t\bar{t}$
at colliders, but with different gauge group -- $SU(2)_W\times U(1)_Y$ in
place of $SU(3)_c$ -- and additional considerations of electroweak
symmetry breaking.

\subsection{Regions Analysis}

\begin{figure}[htb]
\begin{center}
\vspace{5mm} 
\input{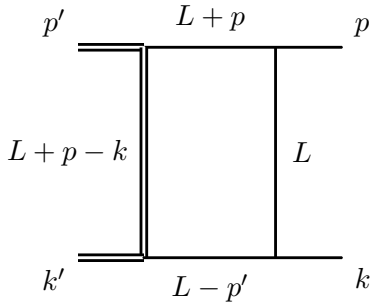}
\vspace{5mm} 
\caption{\label{fig:wino} 
Diagrammatic representation of the integral (\ref{eq:int}). The double lines 
are propagators for a state with mass $M$, while the single lines are appropriate for massless particles.
}
\end{center}
\end{figure}

Different fields in the (soft collinear) effective theory will
correspond to  different  momentum modes for the various particles in
the original theory. To derive the fields required to reproduce the IR structure
of the full theory, we analyze  the singularity structure of diagrams
that contribute to heavy WIMP annihilation.   This systematic
decomposition of loop integrals is  known as a regions analysis
(for a monograph on this subject, see \cite{Smirnov:2002pj}).  It
simultaneously allows  for the perturbative solution of the integrals
when a separation of scales is present, while providing insight as to
what modes are required to construct an EFT that can be
matched to the full theory order-by-order in the gauge coupling and power counting parameter $\lambda=m_W/M$.

For concreteness, let us consider, \emph{e.g.}, the integral
\begin{align}\label{eq:int} 
I &= 
\int (dL) {\frac{1}{ L^2}} {\frac{1}{ (L+p)^2}}{\frac{1}{ (L-p^\prime)^2} }
\frac{1}{ (L + p-k)^2 - M^2 } \,. 
\end{align}
Apart from numerator structure (inessential for the regions analysis), this 
integral
corresponds to the diagram in Fig.~\ref{fig:wino}.  We use the shorthand notation,  
\be 
(dL) = {d^d L \over (2\pi)^d} \,, 
\ee
and employ dimensional regularization with $d=4-2\epsilon$ dimensions. 

The physical process of interest involves initial state heavy particles at rest annihilating to massless energetic 
particles.  It is therefore useful to introduce the timelike unit vector $v^\mu$ with $v^2=1$, 
and lightcone vectors $n^\mu$ and $\bar{n}^\mu$ satisfying 
$n^2=\bar{n}^2=0$ and $n\cdot \bar{n} = 2$.  
For momenta in the $\pm \hat{z}$ direction, a convenient choice is $n^\mu =(1,0,0,1)$, $\bar{n}^\mu=(1,0,0,-1)$.
While allowing a more general relation is convenient for some 
purposes
(such as analyzing Lorentz invariance constraints of subleading corrections as done in \cite{Manohar:2002fd, Hill:2012rh}), for simplicity we take $2v=n+\bar{n}$.   
We take the heavy WIMPs to have momentum
\be
k = k^\prime = Mv \,. 
\ee
For the massless final state particles, it is convenient to expand their momenta in lightcone components, 
\be
p^\mu \quad \longleftrightarrow \quad (n\cdot p, \, \bar{n}\cdot p, \, p_\perp) \,,
\ee
where $p_\perp^\mu = p^\mu - (n\cdot p) \bar{n}^\mu/2 - (\bar{n}\cdot p) 
n^\mu/2$.  
Let us consider the integral representing an
amplitude with (offshell) final state momenta 
$p^2 \sim p^{\prime 2} \sim M^2\lambda$, where $\lambda$ is 
the dimensionless power counting 
expansion parameter of SCET.   For example, writing 
\be
p = M n + \delta p \,, \quad p^\prime = M \bar{n} - \delta p \,, 
\ee
such that $p+p^\prime = 2M v$, we may take 
$\delta p^\mu = \delta p_\perp^\mu$ so that 
$p^2 = p^{\prime 2} = (\delta p)^2 \sim M^2\lambda$.  
Evaluating the integral \eqref{eq:int} in the limit $p^2/M^2 \sim \lambda \ll 1$, we have
\be\label{eq:fullint}
I = {[c_\epsilon] \over M^4} \bigg[ -\frac14 \log^2\left(- p^2 + 
i 0 
\over 4 M^2\right) 
- {\pi^2\over 48} - {i\pi \over 4} \log\left(-p^2 + 
i0 
\over 4 M^2\right) 
+ \order( p^2/M^2 ) \bigg]
\,. 
\ee

Consider the following momentum regions (decomposed along the light cone):
\begin{align}\label{eq:scaling}
(\rm hard) \quad h: &\quad L^\mu \sim M(1,1,1) \,, 
\nl
{(\rm soft)} \quad s: &\quad L^\mu \sim M(\lambda,\lambda,\lambda) \,,
\nl
(\rm hardcollinear) \quad hc: &\quad L^\mu \sim M(\lambda,1,\lambda^{\frac12}) 
\,,
\nl
(\text{anti-hardcollinear}) \quad {\rm \overline{hc}}: &\quad L^\mu \sim 
M(1,\lambda,\lambda^{\frac12}) \,.
\end{align}
Now we will show that these regions are sufficient 
to reproduce the full theory result, to leading order in $\lambda$. Taylor expanding the 
four-momentum $L$ in each denominator of (\ref{eq:int}) following the scalings in (\ref{eq:scaling}) gives the integrals
\begin{align}
\label{eq:decomp1}
I_{\rm h} &= 
\parbox{20mm}{
\begin{fmfgraph*}(50,40)
  \fmfleftn{l}{2}
  \fmfrightn{r}{2}
  \fmf{double}{l2,a}
  \fmf{double}{l1,a}
  \fmf{plain}{a,r2}
  \fmf{plain}{a,r1} 
\end{fmfgraph*}
}
=
\int (dL) {1\over L^2} {1\over L^2 + 2M n\cdot L} {1\over L^2 - 2M 
\bar{n}\cdot L} 
{1\over L^2 + M(n\cdot L - \bar{n}\cdot L) - 2M^2} \,,
\nl
I_s &= 
\parbox{20mm}{
\begin{fmfgraph*}(50,40)
  \fmfleftn{l}{2}
  \fmfrightn{r}{2}
  \fmf{double}{l2,a}
  \fmf{double}{l1,a}
  \fmf{plain}{a,r2}
  \fmf{plain}{a,r1} 
  \fmffreeze
  \fmf{phantom}{a,b,r1} 
  \fmf{phantom}{a,c,r2} 
  \fmffreeze
  \fmf{plain}{b,c}
\end{fmfgraph*}
}
=
\int (dL) {1\over L^2} {1\over p^2 + 2M n\cdot L} {1\over p^{\prime 2} - 
2M \bar{n}\cdot L} {-1\over 2M^2} \,,
\nl
I_{\rm hc} &= 
\parbox{20mm}{
\begin{fmfgraph*}(50,40)
  \fmfleftn{l}{2}
  \fmfrightn{r}{2}
  \fmf{double}{l2,a}
  \fmf{double}{l1,a}
  \fmf{plain}{a,r2}
  \fmf{plain}{a,r1} 
  \fmffreeze
  \fmf{phantom}{a,b,r2} 
  \fmffreeze
  \fmf{plain,right}{a,b}
\end{fmfgraph*}
}
=
\int (dL) {1\over L^2} {1\over (L+p)^2} {1\over -2M \bar{n}\cdot L} 
{1\over -M\bar{n}\cdot L - 2 M^2} \,,
\nl
I_{\overline{\rm hc}} &= 
\parbox{20mm}{
\begin{fmfgraph*}(50,40)
  \fmfleftn{l}{2}
  \fmfrightn{r}{2}
  \fmf{double}{l2,a}
  \fmf{double}{l1,a}
  \fmf{plain}{a,r2}
  \fmf{plain}{a,r1} 
  \fmffreeze
  \fmf{phantom}{a,b,r1} 
  \fmffreeze
  \fmf{plain,left}{a,b}
\end{fmfgraph*}
}
=
\int (dL) {1\over L^2} {1\over 2M n\cdot L} {1\over 
(L-p^\prime)^2} {1\over Mn\cdot L - 2M^2} \,.
\end{align}
There are also contributions from the momentum routing where the lines with $L+p$ and 
$L-p^\prime$ in Fig.~\ref{fig:wino} become soft.  With momentum $L$ for the soft 
line, 
\begin{align}
\label{eq:decomp2}
I_{s,{\rm hc}} &= 
\parbox{20mm}{
\begin{fmfgraph*}(50,40)
  \fmfleftn{l}{2}
  \fmfrightn{r}{2}
  \fmf{double}{l2,a}
  \fmf{double}{l1,a}
  \fmf{plain}{a,r2}
  \fmf{plain}{a,r1} 
  \fmffreeze
  \fmf{phantom}{l2,b,a} 
  \fmf{phantom}{a,c,r2}
  \fmffreeze
  \fmf{plain}{b,c}
\end{fmfgraph*}
}
= \int (dL) {1\over p^2-2M n\cdot L} {1\over L^2} {1\over 4M^2} { 1 
\over -M(n\cdot L + \bar{n}\cdot L)} \,,
\nl
I_{s,\overline{\rm hc}} &=
\parbox{20mm}{
\begin{fmfgraph*}(50,40)
  \fmfleftn{l}{2}
  \fmfrightn{r}{2}
  \fmf{double}{l2,a}
  \fmf{double}{l1,a}
  \fmf{plain}{a,r2}
  \fmf{plain}{a,r1} 
  \fmffreeze
  \fmf{phantom}{l1,b,a} 
  \fmf{phantom}{a,c,r1}
  \fmffreeze
  \fmf{plain}{b,c}
\end{fmfgraph*}
}
= 
\int (dL) {1\over p^{\prime 2} + 2M \bar{n}\cdot L} {1\over 4M^2} {1\over 
L^2}{1\over M (n\cdot L + \bar{n}\cdot L) } \,.
\end{align} 
An explicit evaluation of these integrals yields 
\begin{align}\label{eq:sum}
I_{\rm h} &= M^{-2\epsilon} \left[ -{1\over 4\epsilon^2} + {1\over \epsilon}\left( 
\frac12\log{2} - {i\pi\over 8} \right)
+ {5\pi^2\over 48} - \frac12 \log^22 + {i\pi\over 4}\log{2} \right] 
\,,
\nl
I_s &= \left( - {p^2 p^{\prime 2} \over 4 M^2} + i 0 \right)^{-\epsilon} \left( 
-{1\over 8\epsilon^2} - {\pi^2\over 48} \right) 
\,,
\nl
I_{\rm hc} &= (-p^2-i0)^{-\epsilon} \left( {1\over 4\epsilon^2} - {\pi^2\over 24} 
\right)
\,,
\nl
I_{\overline{\rm hc}} &= (-p^{\prime 2} - i0)^{-\epsilon} \left( {1\over 
4\epsilon^2} - {\pi^2\over 24} \right) 
\,,
\nl
I_{s,{\rm hc}} &= \left(- {p^2\over 2M} - i0 \right)^{-2\epsilon} \left( -{1\over 16 
\epsilon^2} - {\pi^2\over 24} \right)
\,,
\nl
I_{s,\overline{\rm hc}} &= \left( -{p^{\prime 2}\over 2M} - i0 \right)^{-2\epsilon} 
\left( -{1\over 16\epsilon^2} - {\pi^2\over 24} \right)
\,,
\end{align} 
where overall factors of
$M^{-4}$ and $[c_\epsilon ] = i(4\pi)^{-2+\epsilon} \Gamma(1+\epsilon)$ have been dropped
for simplicity.  One can verify that the integrals in (\ref{eq:sum}) 
sum to the expression (\ref{eq:fullint}) for the 
original integral (\ref{eq:int}). 

This demonstrates the field content required for a complete EFT description of the diagram in Fig.~\ref{fig:wino}. The
hard scale $L^\mu \sim M(1,1,1)$ will be captured entirely in the Wilson coefficient of the annihilation operator 
in the EFT through matching at the scale $\mu \sim 2 M$.  Then the contributions from IR modes (at leading power $\lambda^0$) 
are reproduced by momentum scalings that we identify with ``hardcollinear" and ``soft" regions.  A Lagrangian field theory 
with fields corresponding to these modes can be constructed.  The  Feynman rules of this EFT will correspond to the explicit classes 
of diagrams of (\ref{eq:decomp1}) and (\ref{eq:decomp2}), thereby encoding the entire IR structure of the full theory as 
a controlled expansion in an explicit small parameter.

A similar analysis can be used to demonstrate that other diagrams are 
reproduced in the same manner by the sum over momentum regions
of dimensionally regulated integrals. For the matching at the weak scale $\mu \sim m_W$, we may perform a similar analysis to isolate contributions from the potential region contained in the soft region of the diagrams. The potential region has scaling $v \cdot L \sim M \lambda^2$ and $(L^\mu - v^\mu v \cdot L)  \sim M \lambda$, and is resummed as the Sommerfeld enhancement by solving the quantum mechanical Hamiltonian.

This procedure allows for the systematic factorization of momentum regions: the scale $M$ will \emph{only} appear in the hard matching coefficient (up to the collinear anomaly discussed in Sec.~\ref{sec:CollinearAnomaly} below), and the EFT will only depend on the IR scale $m_W$. By evolving the Wilson coefficients from $\mu \sim 2M$ to $m_W$ using the RGEs of the EFT, the large logarithms discussed above are resummed, thereby systematically improving perturbation theory. The rest of this section provides the explicit construction of this EFT for application to heavy WIMP annihilation.

\subsection{Heavy Particle and Soft Collinear Effective Theory for WIMP 
Annihilation\label{sec:SCET}}

Having motivated the introduction of soft, hardcollinear and 
anti-hardcollinear modes, we now proceed to construct an effective 
theory describing interactions at scales $m_W^2 \ll \mu^2 \ll M^2$.  
We perform this analysis in the electroweak symmetric vacuum; accounting for the effects of 
electroweak 
symmetry breaking will be discussed in Sec.~\ref{sec:lowmatch} below. 

We focus for simplicity on a self-conjugate scalar WIMP, necessarily 
a $U(1)_Y$ hypercharge singlet that transforms under an integer isospin 
representation 
of $SU(2)_W$.   
We ignore Standard Model field content beyond the $SU(2)_W$ gauge 
fields; modifications to this case are straightforward.  
In the absence of collinear degrees of freedom, the heavy WIMP is
described as a heavy particle field, with Lagrangian~\cite{Hill:2011be}
\be
{\cal L}_{\phi_v} = \phi_v^* ( iv\cdot D + \dots ) \phi_v \,,
\ee
where $\phi_v$ denotes the scalar heavy particle field, $v^\mu$ is the heavy 
particle velocity introduced above, and $D_\mu$ is the $SU(2)_W$ covariant
derivative (\ref{eq:covderiv}). 

The soft, hardcollinear and anti-hardcollinear gauge fields are 
denoted by $A_{s}^\mu$, $A_{\rm hc}^\mu$ and $A_{\overline{\rm hc}}^\mu$, 
and are described by 
respective Lagrangians that are formally identical to those for 
the full $SU(2)_W$ gauge theory, with the understanding that each field 
is restricted to the appropriate momentum mode.  
We suppress the matrix structure, $A_\mu \equiv A_\mu^a T^a$, and, to avoid conflicting notation with Wilson lines below, denote
the $SU(2)_W$ gauge field by $A_\mu$ (instead of $W_\mu$).  
Corresponding to the scalings in (\ref{eq:scaling}), a power counting 
in which the gauge field components scale in the same way as their momentum
is assigned: 
\be
A_{s}^\mu \sim (\lambda,\lambda,\lambda), \quad
A_{\rm hc}^\mu \sim (\lambda,1,\lambda^{\frac12}), \quad \text{and}\quad
A_{\overline{\rm hc}}^\mu \sim (1,\lambda,\lambda^{\frac12}).
\ee
In this way, 
Lagrangian interactions may be expanded as a series in $\lambda$.  Amputated Feynman diagrams and corresponding $S$ matrix elements will
obey a simple power counting based on the appearance
of the associated 
vertices~\cite{Bauer:2000ew,Bauer:2000yr,Chay:2002vy,Beneke:2002ph,Hill:2002vw}.  Gauge fixing and ghosts can be treated in the standard way.

This power counting implies that leading order 
interactions may occur between soft and hardcollinear fields 
(or between soft and anti-hardcollinear fields), 
since, \emph{e.g.} $n\cdot A_s \sim n\cdot A_{\rm hc} \sim \lambda$.   At leading order
the interactions of the soft field with the hardcollinear sector
are given by the replacement in the hardcollinear Lagrangian,
\be\label{eq:hcreplace}
A_{\rm hc}^\mu(x) \to A_{\rm hc}^\mu(x) + n\cdot A_s(x_-) {\bar{n}^\mu\over 2}
\,,
\ee
where $x_-^\mu \equiv (\bar{n}\cdot x) n^\mu/2$ and  
$x_+^\mu \equiv (n\cdot x) \bar{n}^\mu/2$ are arbitrary four-vectors expanded along the light cone.  The ``multipole'' expansion 
of $A_s(x) = A_s(x_-) + \order(\lambda^2)$ ensures that only the $n\cdot p_s$ 
components of soft momenta
are added to hardcollinear momenta.  Similar considerations, with 
$n\leftrightarrow \bar{n}$, 
apply to the interactions between soft and anti-hardcollinear fields. 

The local gauge invariance of the full theory is mapped to 
separate soft, hardcollinear and anti-hardcollinear gauge transformations in
the effective theory, 
\begin{align}
s: &\quad 
A_{\rm hc}^\mu \to V_s (x_-) A_{\rm hc}^\mu V_s^\dagger(x_-) \,,
\quad 
A_{\overline{\rm hc}}^\mu \to V_s (x_+) A_{\rm hc}^\mu V_s^\dagger(x_+) \,,
\quad 
A_s^\mu \to V_s  A_s^\mu V_s^\dagger + {i\over g} V_s \big[ \partial^\mu 
,V_s^\dagger \big] \,,
\nl
{\rm hc}: 
&\quad 
A_{\rm hc}^\mu \to V_{\rm hc}  A_{\rm hc}^\mu V_{\rm hc}^\dagger
+  {i\over g} V_{\rm hc} \big[ \partial^\mu - i g A_{s,+}^\mu(x_-) , V_{\rm 
hc}^\dagger \big] \,,
\quad 
A_{\overline{\rm hc}}^\mu \to A_{\overline{\rm hc}}^\mu \,,
\quad 
A_s^\mu \to A_s^\mu  \,,
\nl
\overline{{\rm hc}}: 
&\quad 
A_{{\rm hc}}^\mu \to A_{{\rm hc}}^\mu \,,
\quad 
A_{\overline{\rm hc}}^\mu \to V_{\overline{\rm hc}}  A_{{\overline {\rm hc}}}^\mu 
V_{\overline{\rm hc}}^\dagger
+  {i\over g} V_{\overline{\rm hc}} \big[ \partial^\mu - i g A_{s,-}^\mu(x_+) , 
V_{\overline{\rm hc}}^\dagger \big] \,,
\quad 
A_s^\mu \to A_s^\mu  \,.  
\end{align}

With these preliminaries, we can determine the leading order basis of operators representing
heavy WIMP annihilation to di-boson final states.
Since components of the derivatives and gauge fields count as $\order(1)$ 
in the power counting, \emph{e.g.}, $\bar{n}\cdot A_{\rm hc} \sim 1$, operators are built from
field combinations that implement lightcone gauges 
$\bar{n}\cdot {\cal A}_{\rm hc} = n\cdot {\cal A}_{\overline{\rm hc}}=0$.
Expressed in an arbitrary gauge, these fields read, 
\be
g{\cal A}_{\rm hc}^{\mu} = W^\dagger iD_{\rm hc}^{\mu} W \,, 
\quad
g{\cal A}_{\overline{\rm hc}}^{\mu} = \overline{W}^\dagger iD_{\overline{\rm hc}}^{\mu} 
\overline{W} \,,
\ee
where $iD_{{\rm hc}(\overline{\rm hc})}^\mu = i\partial^\mu + g ( A_{\rm hc 
(\overline{\rm hc})}^\mu + A_{s\pm }(x_\mp)^\mu )$, 
and $W$ ($\overline{W}$) is a Wilson line of hardcollinear
(anti-hardcollinear) fields in the $\bar{n}$ ($n$) direction, 
\be
W(x) = {\rm P}\exp\bigg[ ig\int_{-\infty}^0 ds\, \bar{n}\cdot A_{\rm hc}(x+s\,\bar{n}) 
\bigg] \,,
\quad
\overline{W}(x) = {\rm P}\exp\bigg[ ig\int_{-\infty}^0 ds \, n\cdot 
A_{\overline{\rm hc}}(x+s\,n) \bigg] \,. 
\ee
Noting the scaling relations, 
\begin{align}
{\cal A}_{\rm hc}^\mu \sim (\lambda,0, \lambda^{\frac12}) \,, \quad
{\cal A}_{\overline{ \rm hc}}^\mu \sim (0, \lambda, \lambda^{\frac12}) \,, 
\end{align}
we see that operators mediating leading-order processes with two 
initial state heavy WIMPs, one final state hardcollinear field and 
one final state anti-hardcollinear field are of the form
$\phi_v^a \phi_v^b {\cal A}_{\rm hc\,\perp}^{c\,\mu} {\cal A}_{\overline{\rm hc}\,\perp}^{d\,\nu}$, 
with gauge indices $a,b,c,d$ contracted to form invariant combinations.
It is straightforward to see that, for an arbitrary $SU(2)_W$ representation, there are two such operators,
\be\label{eq:Leff}
{\cal L}_{\phi_v\phi_v} = {1\over M} \sum_{i=1}^2 c_{i}\, O_i + {\rm h.c.} + 
\order(1/M^2) \,,
\ee
where the explicit form of these dimension 5 operators is given by
\begin{align}
\label{eq:originalbasis}
O_1 &= g^2\, \phi_v^T \, \phi_v \, {\cal A}^a_{{\rm hc}\,\perp\, \mu} \, {{\cal A}^a_{{\rm \overline{hc}}\,\perp}}^\mu 
=  g^2 \, \phi_v^i \, \phi_v^j \, {\cal A}_{{\rm hc}\, \mu}^a \, {\cal A}^b_{{\rm \overline{hc}}\, \nu} \, \delta^{ij} \,\delta^{ab} \, g_\perp^{\mu\nu} \,,
\nl
O_2 &= 
g^2 \, \phi_v^T \, {\cal A}_{\rm{hc}\,\perp\, \mu} \, {{\cal A}_{\rm{\overline{hc}}\,\perp}}^\mu \,\phi_v 
=  g^2 \, \phi_v^i \, \phi_v^j \, {\cal A}_{\rm{hc}\, \mu}^a \, {\cal A}^b_{\rm{\overline{hc}}\, \nu} \,(t^a t^b)^{ij} \, g_\perp^{\mu\nu} \,.
\end{align}
Here $g_{\perp}^{\mu\nu} = g^{\mu\nu} - (n^\mu \bar{n}^\nu + \bar{n}^\mu n^\nu)/4$ projects onto transverse components. 
Note that the coupling factor $g^2$ is included in the operator definition 
(as opposed to being absorbed into $c_i$) for convenience in the renormalization analysis.

In the following, we consider the matching of full theory amplitudes at the hard 
scale $\mu \sim 2 M$, and evolve the resulting matching coefficients to the scale $\mu \sim 
m_W$ by computing the anomalous dimensions and solving the evolution equation. 

\subsection{Electroweak Symmetric SCET Feynman Rules}
\label{sec:EWSymSCETFeynRules}

In this section, we give the Feynman rules for the effective theory describing interactions at scales $m_W^2 \ll \mu^2 \ll M^2$.
Note that the presence of the hardcollinear gauge boson operators $\mathcal{A}$ 
in the definition of the $O_i$ implies that there can be an arbitrary
number of gauge boson emissions from the operator vertex insertion.
We use `t Hooft-Feynman gauge in the following. 
The Feynman rules for operator insertions of $O_{1,2}$ are: 
  \\
\begin{align}\label{eq:Ofeynrule}
\qquad
  \parbox{40mm}{
    \begin{fmfgraph*}(100,80)
      \fmfleftn{l}{2}
      \fmfrightn{r}{2}
      \fmf{double}{l2,a}
      \fmf{double}{l1,a}
      \fmf{curly}{a,r2}
      \fmf{curly}{a,r1}
      \fmffreeze
      \fmf{plain}{a,r1}
      \fmf{plain}{a,r2}
      \fmflabel{$p,a,\mu$}{r2}
      \fmflabel{$p^\prime,b,\nu$}{r1} 
      \fmflabel{$M v + k,i$}{l2}
      \fmflabel{$M v + k^\prime,j$}{l1}
      \fmfv{label=$O_m$,label.angle=180,label.dist=.15w,decor.shape=circle,
decor.filled=empty,decor.size=.1w}{a}
    \end{fmfgraph*} }
\hspace{-10mm}
& \quad = 
g^2( T_m^{ab} + T_m^{ba} )_{ij} 
g_\perp^{\mu\nu}\,,
\end{align}
\\
where the color structures, defined as
\be
(T_1^{ab})_{ij} = \delta^{ab}\delta_{ij} \,, \quad
(T_2^{ab})_{ij} = (t^a t^b )_{ij} \, ,
\ee
are taken from (\ref{eq:originalbasis}).  Note that this involves one hardcollinear (top of diagram) and one anti-hardcollinear (bottom of diagram) particle.  We need also the Feynman rule with an additional hardcollinear, or anti-hardcollinear gauge boson from the operator vertex.  The Feynman rule for two hardcollinear and one anti-hardcollinear emissions is
\begin{align}
 \parbox[c][45mm][c]{40mm}{
    \begin{fmfgraph*}(100,80)
      \fmfleftn{l}{2}
      \fmfrightn{r}{4}
      \fmf{double}{l2,a}
      \fmf{double}{l1,a}
      \fmf{curly}{a,r4}
      \fmf{curly}{a,r1}
      \fmffreeze
      \fmf{curly}{a,r3}
      \fmf{plain}{a,r4}
      \fmf{plain}{a,r1}
      \fmf{plain}{a,r3}
      \fmflabel{$p,c,\nu$}{r4}
      \fmflabel{$q,d,\rho$}{r3}
      \fmflabel{$p^\prime,a, \mu$}{r1} 
      \fmflabel{$M v + k,i$}{l2}
      \fmflabel{$M v + k^\prime,j$}{l1}
        \fmfv{label=$O_m$,label.angle=180,label.dist=.15w,decor.shape=circle,
              decor.filled=empty,decor.size=.1w}{a}
    \end{fmfgraph*}
  } &\quad= -2  g^3 (T_m^{ba})_{ij}\,f^{bcd} \left( g_\perp^{\mu\nu} {\bar{n}^\rho \over \bar{n}\cdot q} - 
g_{\perp}^{\mu\rho} {\bar{n}^\nu\over \bar{n}\cdot p } \right)\,.
\end{align}
\\
Similar expressions, with $n\leftrightarrow \bar{n}$, hold for one hardcollinear and two anti-hardcollinear emissions.  
The three- and four-point vertices involving all hardcollinear or all anti-hardcollinear 
gauge bosons are identical to the usual QCD results.  As in (\ref{eq:hcreplace}), 
the leading order interaction of soft gauge bosons with 
hardcollinear gauge bosons is given by the multipole expansion in powers of $\lambda$ of 
\be
{\cal L}_{{\rm hc},\,s} = {g \over 2} f^{abc} n\cdot A_s^c A_{{\rm hc}\,\mu}^a 
( 2\partial^\mu \bar{n}\cdot A_{\rm hc}^b - \bar{n}\cdot \partial A_{\rm hc}^{b\,\mu} ) 
+ \dots \,,
\ee
yielding the Feynman rule (here all momenta are ingoing)
\\
\begin{align}
  \parbox{50mm}{
    \begin{fmfgraph*}(100,60)
      \fmfleftn{l}{3}
      \fmfrightn{r}{3}
      \fmftopn{t}{3}
      \fmf{curly}{l1,a}
      \fmf{curly}{r1,a}
      \fmf{curly}{t2,a}
      \fmffreeze
      \fmf{plain}{a,r1}
      \fmf{plain}{a,l1}
      \fmflabel{$p,a,\nu$}{l1}
      \fmflabel{$k,b,\rho$}{r1}
      \fmfv{label=$q,,c,,\mu$,label.angle=90}{t2}
    \end{fmfgraph*}}
&= g f^{abc} \bar{n}\cdot p \, n^\mu  g_{\perp}^{\nu\rho} + 
\order(\lambda^{\frac12}) \,.
\end{align}
\\
The interaction of soft gauge bosons with heavy scalars is given by the usual result, 
\begin{align}
\nonumber
\\
  \parbox{50mm}{
    \begin{fmfgraph*}(100,60)
      \fmfleftn{l}{3}
      \fmfrightn{r}{3}
      \fmftopn{t}{3}
      \fmf{double}{l1,a}
      \fmf{double}{r1,a}
      \fmffreeze
      \fmf{curly}{t2,a}
      \fmflabel{$i$}{l1}
      \fmflabel{$j$}{r1}
      \fmfv{label=$a,,\mu$,label.angle=90}{t2}
    \end{fmfgraph*}
  }
&= ig (t^a)_{ji} v^\mu\,.
\end{align}
\\

Armed with these Feynman rules, the renormalized Wilson coefficients
$c_i(\mu)$ can be computed by matching the full theory onto the EFT at
$\mu \sim 2 M$ (the subject of Sec.~\ref{sec:hardmatching}).
Furthermore, anomalous dimensions for the operators $O_i$ can be
computed, which determine the RGEs that allow us to compute
$c_i(m_W)$ using $c_i(2M)$ as input (the subject of
Sec.~\ref{sec:RG}).

\section{High Scale Matching \label{sec:hardmatching}}
This section provides the matching calculation between the electroweak 
symmetric full and effective theories at renormalization scale $\mu \sim 2 M \gg m_W$.  Consider the process $\phi_i(k) + \phi_j(k^\prime) \to A^a(p) + A^b(p^\prime)$.  
Given two initial state WIMPs at zero velocity $k=k^\prime=Mv$, conservation of momentum 
implies that 
the massless final state gauge bosons have $p=Mn$ and $p^\prime = M\bar{n}$.  
Therefore, all factors of $\bar{n}\cdot p$ and $n\cdot p^\prime$ will be replaced with $2M$ in what follows.  

\subsection{Matching Conditions}
The matching condition can be stated as 
\begin{align}\label{eq:hardmatch}
{1\over 2M} \bigg(Z_{\phi,{\rm full}}^\frac12\bigg)^2 \bigg(Z_{A,{\rm full}}^\frac12\bigg)^2 \sum_i 
{\cal M}_{i, \rm full}
{ \langle O_i\rangle^{\rm tree} \over g^2(\mu) } 
&= \bigg(Z_{\phi_v}^\frac12 \bigg)^2 \bigg(Z_{A,{\rm eff}}^\frac12\bigg)^2 \sum_i c_i^{\rm bare} 
\langle O_i^{\rm bare}\rangle, 
\end{align}
where the onshell wavefunction factors for the external particles ensure that we are comparing two physical amplitudes
({\`a} la LSZ reduction).  
The factor of $\langle O_i\rangle^{\rm tree}$ on the left hand side accounts for color and polarization structures 
(see (\ref{eq:Ofeynrule}) for the explicit expression).
We have here defined the tree-level matrix element without gauge coupling as 
\be
\langle O_i \rangle^{\rm tree} \equiv {g^2(\mu) \over g_{\rm bare}^2 } \langle O_i^{\rm bare} \rangle^{\rm tree} \,. 
\ee
We will solve this equation for the bare Wilson coefficients $c_i^\text{bare}$.

Since we are working with electroweak symmetric SCET, there are no dimensionful parameters in the theory.  Noting that scaleless integrals are zero 
in dimensional regularization, the effective theory loop integrals and renormalization factors vanish. Hence,
\begin{align}
\bigg(Z_{\phi_v}^\frac12 \bigg)^2 \bigg(Z_{A,{\rm eff}}^\frac12\bigg)^2 \sum_i c_i^{\rm bare} 
\langle O_i^{\rm bare}\rangle 
&= 
Z_g^2 \mu^{2\epsilon} \sum_i c_i^{\rm bare} 
\langle O_i\rangle^{\rm tree}.
\end{align}
It is straightforward to identify the bare matching coefficients with the corresponding
full theory diagrams using (\ref{eq:hardmatch}).

\begin{figure}
\begin{center} 
\input{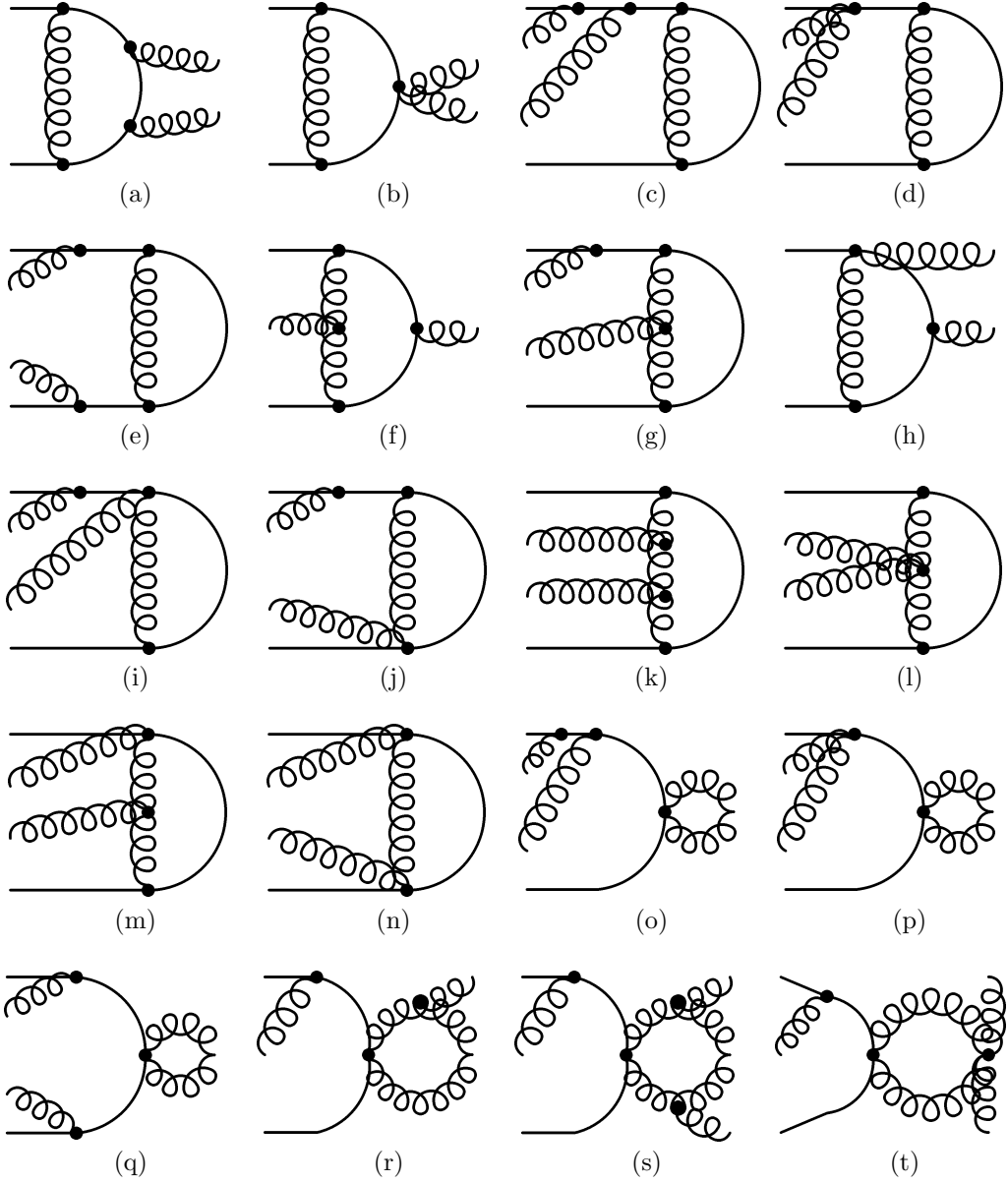}
\caption{Diagrams contributing to hard scale matching. 
}
\label{fig:harddiagrams} 
\end{center}
\end{figure} 

The amputated full theory diagrams are depicted in Fig.~\ref{fig:harddiagrams}.  Note that real emissions from the initial state heavy WIMPs, and the associated vertex corrections, are power suppressed. Real emissions from the final state bosons are relevant for the $W^+\,W^-$ annihilation channel, and is left to future work~\cite{future}.
In terms of the bare coupling constant $\tilde{g}_{\rm bare}$ of the full theory, the 
resulting amplitudes 
read, 
\begin{align}\label{eq:Miresults}
{\cal M}_{1,{\rm full}} &= -i[c_\epsilon] \tilde{g}_{\rm bare}^4 
(2M)^{-2\epsilon}\bigg\{
 {1\over \epsilon} ( 2 - 2i\pi )C_2(j) 
+ C_2(j)\left( - 1 - {\pi^2\over 4} \right)  
\bigg\}
\,,
\nl
{\cal M}_{2,{\rm full}} &= {\tilde{g}_{\rm bare}^2}
- i [c_\epsilon] \tilde{g}_{\rm bare}^4 (2M)^{-2\epsilon}\bigg\{ 
-{4\over \epsilon^2} 
+{1\over \epsilon} \bigg[ -6 + 2i\pi \bigg]
+ C_2(j) \left( -7 - {\pi^2\over 4} \right) + 7 + {29\pi^2\over 12} 
\bigg\} \,,  
\end{align} 
where $C_2(j) = j(j+1)$ is the quadratic Casimir invariant for the spin-$j$ 
representation of $SU(2)_W$.  Note that we distinguish $\tilde{g}$ and $\tilde{Z}_g$ in the full theory from $g$ and $Z_g$ in the effective theory, 
which differ because the heavy WIMP has been integrated out below the scale $M$ and as such no longer contributes to the running of the gauge coupling. 
Specifically, at one loop the relation between $g$ and $\tilde{g}$ is \cite{Pierce:1996zz}
\be
\label{eq:gDecoupling}
{\tilde{g}^2 \over g^2} = 1 - { C(j) \over 3 }{g^2\over (4\pi)^2} \log { M \over \mu} \,.
\ee
Working in `t Hooft-Feynman gauge, the onshell wavefunction factors for the full theory fields in the electroweak symmetric vacuum can be derived at one loop to be
\begin{align}\label{eq:Zfactors}
Z_{\phi,{\rm full}} &=  1 \,,
\nl 
Z_{A,{\rm full}} &= 1 + {{\tilde g}^2\over (4\pi)^2}\left(\frac{M}{\mu}\right)^{-2\epsilon}\bigg[ -{1\over 6\epsilon} C(j) + 
\order(\epsilon) \bigg] \,,
\end{align}
where ${\rm tr}(t^ct^d) \equiv C(j) \delta^{cd}$, so that $C(j) = 
j(j+1)(2j+1)/3$.  Note that only the heavy WIMP contributes to $Z_{A,{\rm full}}$ since the Standard Model matter is massless and therefore the corresponding integrals are zero in dimensional regularization. 
In the final result for renormalized hard coefficients, the finite term in $Z_{A,{\rm full}}$ cancels with the contribution from the decoupling relation in (\ref{eq:gDecoupling}).  
To relate $\tilde{g}^{\rm bare}$ in (\ref{eq:Miresults}) to $\tilde{g}(\mu)$ and hence $g(\mu)$ in (\ref{eq:gDecoupling}), 
we require 
\be
\label{eq:gMSbarFull}
\tilde{g}_{\rm bare} = \tilde{Z}_g \mu^\epsilon \tilde{g}(\mu) \,, \quad
\tilde{Z}_g = 1 + {\tilde{g}^2\over (4\pi)^2}{1\over \epsilon}\bigg[ 
{1\over 12} C(j) 
-{43\over 12} + \frac23 n_G 
\bigg] \,, 
\ee
where we have included $n_G=3$ generations of Standard 
Model fermions, the Standard Model Higgs doublet, and the heavy scalar WIMP contributions.\footnote{
The one-loop correction is proportional to $\frac53 C_2(j=1) - \frac23 [n_G (N_c 
+1)]C(j=1/2) -\frac13 C(j=1/2)$, 
where the three terms correspond to $SU(2)_W$ gauge bosons, Standard Model 
fermions with 
$n_G=3$ generations and $N_c=3$ colors, and a Higgs doublet respectively.  
The $C(j)/12$ term accounts for the scalar WIMP contribution.
}

The bare coefficients $c_i^{\rm bare}$ are obtained from (\ref{eq:hardmatch}), employing the results (\ref{eq:Miresults}) and 
(\ref{eq:Zfactors}) for the full theory side of the matching condition.  
In the next section, we determine the counterterms in the EFT, such that the renormalized coefficients $c_i(\mu)$ of the effective Lagrangian (\ref{eq:Leff}) can be derived and used as input to the RGEs.
\subsection{Renormalized Matching Coefficients}

\begin{figure}
\input{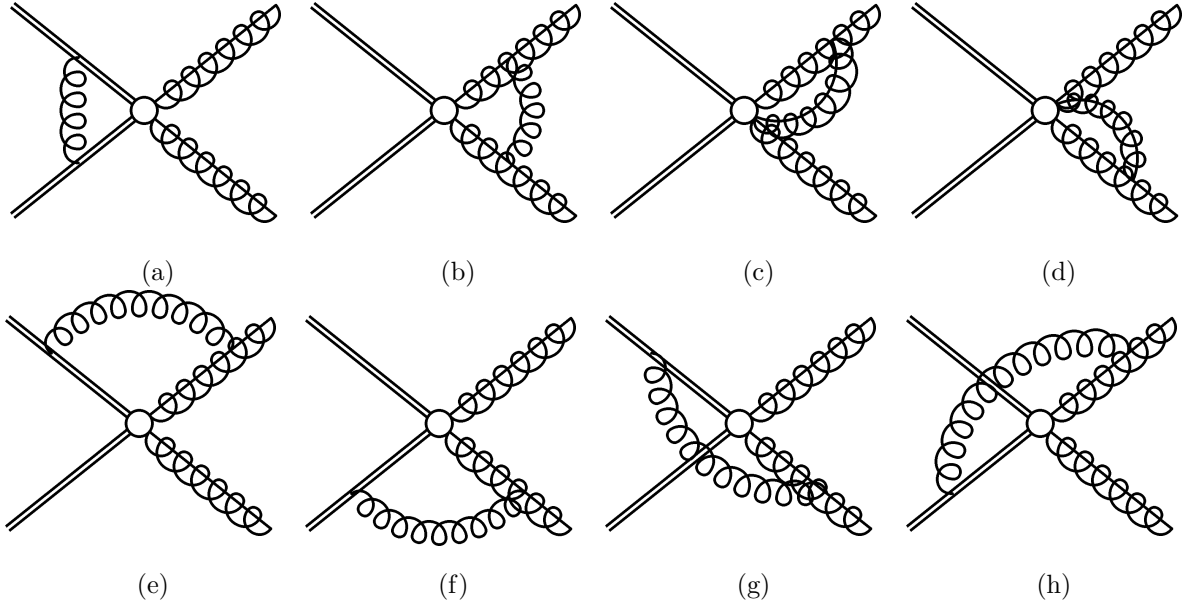}
\caption{One-loop contributions to matrix elements of $O_i$.  
\label{fig:scetoneloop}}
\end{center}
\end{figure} 

In this section, we renormalize $c_i^{\rm bare}$.  Given the Feynman rules for electroweak symmetric SCET provided in Sec.~\ref{sec:EWSymSCETFeynRules} above, 
we can compute one-loop contributions to the matrix elements of $O_i$ between external states 
of two scalars and two hardcollinear gauge bosons.  We regulate infrared divergences with offshell momenta $p$ and $p^\prime$ for the final state 
gauge bosons and $k$ and $k^\prime$ for the initial state heavy WIMPs.  Note that this regulation introduces a scale, allowing us access to the UV 
divergent contributions to the one-loop diagrams, which in turn can be used to derive the counterterms needed to renormalize the theory.  

We will need expressions for the renormalized gauge coupling, and wavefunction renormalization factors 
for the heavy scalar and gauge bosons in the EFT.  
Employing `t Hooft Feynman gauge, the one-loop expressions are 
\be
\label{eq:gMSbarEFT}
g_{\rm bare} = Z_g \mu^\epsilon g(\mu) \,, \quad
Z_g = 1 + {g^2\over (4\pi)^2}{1\over \epsilon}\bigg[
-{43\over 12} + \frac23 n_G 
\bigg] \,, 
\ee
and
\begin{align}\label{eq:ZAeff}
Z_{\phi_v,\text{EFT}} &= 1 + {g^2\over (4\pi)^2}{1\over \epsilon}\left[ 2 C_2(r) \right] 
\,,
\quad 
Z_{A,\text{EFT}} = 1 + {g^2 \over (4\pi)^2} {1\over \epsilon} \left[ 
\frac{19}{6} - \frac43 n_G 
\right] 
\,,
\end{align}
where in $Z_A$ we account for $n_G=3$ generations of Standard Model fermions, and 
the Standard Model Higgs doublet as in (\ref{eq:gMSbarFull}) above.  

Evaluating the diagrams in Fig.~\ref{fig:scetoneloop} yields the UV divergences of the effective theory matrix elements.  Multiplying by appropriate $Z$ factors to obtain physical $S$ matrix elements yields
\begin{align}
Z_{\phi_v,\text{EFT}} Z_{A,\text{EFT}} \langle O_1^{\rm bare} \rangle  
&= \langle O_1 \rangle^{\rm tree} Z_g^2 \bigg\{ 
1 + {g^2\over (4\pi)^2}\bigg[ {4\over \epsilon^2}  + {1\over \epsilon}\left( 
{43\over 6} - {4 \over 3}n_G 
 - 4\log{4M^2\over \mu^2} + 4i\pi\right)
\bigg] \bigg\}
\,,
\nl
Z_{\phi_v,\text{EFT}} Z_{A,\text{EFT}} \langle O_2^\text{bare} \rangle 
&= 
\langle O_1 \rangle^{\rm tree} Z_g^2 {g^2\over (4\pi)^2}{1\over \epsilon}\left[ -2 
C_2(r) + 2i\pi C_2(r) \right] 
\nl
&\quad 
+ \langle O_2 \rangle^{\rm tree} Z_g^2 \bigg\{ 
1+ {g^2\over (4\pi)^2}\bigg[ {4\over \epsilon^2}  + {1\over \epsilon}\left( 
{79\over 6}- {4 \over 3}n_G
 - 4\log{4 M^2\over \mu^2} -2i\pi\right)
\bigg] \bigg\} \,,
\end{align} 
from which we read off the operator renormalization matrix, 
\be
\vec{O}^{\rm bare} = \hat{Z} \vec{O}(\mu) \,,
\ee
where it is understood that $g^{\rm bare}$ in $O_i^{\rm bare}$ is 
expressed in terms of renormalized $g(\mu)$, and the factors 
$Z_g^2\mu^{2\epsilon}$ are absorbed into $\hat{Z}$.  
The entries of the operator renormalization matrix are thus
\begin{align}\label{eq:opZ}
Z_{11} &= 1 + {g^2\over (4\pi)^2}\bigg[ {4\over \epsilon^2}  + {1\over 
\epsilon}\left( 
 - 4\log{4M^2\over \mu^2} + 4i\pi\right)
\bigg] \,,
\nl
Z_{12} &= 0 \,,
\nl
Z_{21} &=  {g^2\over (4\pi)^2}{1\over \epsilon}\left[ -2 C_2(r) + 2i\pi C_2(r) 
\right] \,,
\nl
Z_{22} &= 1+ {g^2\over (4\pi)^2}\bigg[ {4\over \epsilon^2}  + {1\over 
\epsilon}\left( 
6  - 4\log{4 M^2\over \mu^2} -2i\pi\right)
\bigg] \,,
\end{align}
where, as mentioned above, we have set $\bar{n}\cdot p = n\cdot p^\prime = 2M$ appropriate for the kinematics of interest.  

Now we have all the ingredients necessary to derive the renormalized Wilson coefficients $c_i(\mu)$.  Combining the expression for $c_i^{\rm bare}$ derived from (\ref{eq:hardmatch}) with the result for operator renormalization (\ref{eq:opZ}), we obtain 
\be\label{eq:matching}
\vec{c}(\mu) =  \hat{Z}^T \vec{c}^{\,\, \rm bare}  \,.
\ee
Expressed in terms of the renormalized gauge coupling, the renormalized 
Wilson coefficients are given by
\begin{empheq}[box={\mybluebox[5pt]}]{align}
\label{eq:renormc}
c_1(\mu) &= {1\over 2}{g^2\over (4\pi)^2} 
\bigg[ - (4-4\pi i ) C_2(r) \log{2M\over \mu} + C_2(r)\left(- 1 - 
{\pi^2\over 4} \right) \bigg]
+ \order(g^4) 
\,,
\nl
c_2(\mu) &=  {1\over 2}\bigg\{ 1 + {g^2\over (4\pi)^2}\bigg[ -8 \log^2{2M\over \mu} + (12 - 4\pi i) \log{2M\over 
\mu} 
\nl
&\quad 
+ \left(-7-{\pi^2\over 4} \right) C_2(r) + 7 + {29\pi^2\over 12} \bigg]
\bigg\}
+ \order(g^4) 
\,.
\end{empheq}
These are the hard scale matching coefficients.  
In the next section, the RGEs will be derived 
to evolve 
these coefficients down to the weak scale.

\section{Renormalization Group Evolution \label{sec:RG}}
Robust predictions of the annihilation cross section for heavy WIMPs
demand control over the Sudakov-type logarithms, \emph{e.g.}, appearing at
$\order(e^2 g_2^2)$ in the  amplitude ${\cal M}^{+- \to \gamma \gamma}$
given in \eqref{eq:onestep1PICharged}. In this section, we investigate
the resummation of such large contributions by solving the evolution
of the coefficients $c_i(\mu)$ appearing in (\ref{eq:Leff}) from
the hard annihilation scale $\mu_H \sim 2M$ down to the  electroweak scale
$\mu_L \sim m_W$.  The anomalous dimension for the basis
of operators in \eqref{eq:originalbasis} follows from renormalization properties of Wilson lines, and is given by an ansatz for
the anomalous dimension of $n$-jet operators in SCET~\cite{Beneke:2009rj,Becher:2003kh,Becher:2009qa,Becher:2009kw}.
We illustrate the explicit connection between the universal cusp piece
and the Sudakov double log, and present ingredients necessary for
resummation through leading log (LL) and next-to-leading log (NLL) accuracy.

\subsection{Anomalous Dimensions}
The scale evolution of coefficients is governed by the RGE
\begin{align}\label{eq:RGE1}
 \frac{d}{d\log\mu}\,\vec{c}(\mu)=\hat \Gamma^T \,\vec{c}(\mu)\,, \quad \hat\Gamma
&= \hat{Z}^{-1} {d\over d\log\mu} \hat{Z} \, ,
\end{align}
where $\hat \Gamma$ denotes the anomalous dimension.
With ${\hat Z}$ given in \eqref{eq:opZ}, we obtain
\begin{align}\label{eq:Gresult}
{\hat \Gamma} &= {g^2\over (4\pi)^2} \left(\begin{array}{cc} 
8\log{4M^2 \over \mu^2} 
- 8i\pi & 0 \\
4C_2(j) - 4i \pi C_2(j) & \ \ 8\log{4M^2\over \mu^2} 
-12
+ 4i\pi 
\end{array}
\right)  + \order(g^4).
\end{align}
The logarithmic scaling of the diagonal elements is a universal feature related to the cusp anomalous dimension of Wilson loops, which can be identified as the 
origin of the large Sudakov logarithm in \eqref{eq:onestep1PICharged}. The non-cusp part of the anomalous dimension depends on the 
gauge representations of the external states. 
It is convenient to rotate to a basis of 
operators with definite isospin $R=0$ and $R=2$, given respectively by $O_1^\prime = O_1$ and $O_2^\prime = 2O_1/3 - O_2$.  In this basis the anomalous dimension is diagonal,
\begin{align}\label{eq:Gisospin}
{\hat \Gamma}^\prime
= {g^2\over (4\pi)^2} \left(\begin{array}{cc} 
8\log{4M^2\over \mu^2} 
- 8i\pi & 0 \\
0 & 8\log{4M^2\over \mu^2} 
-12
+ 4i\pi 
\end{array}\right) 
+ \order(g^4) 
\,.
\end{align}
We may then identify ${\hat \Gamma}^\prime$ with an ansatz for the anomalous 
dimension of an operator describing a particle of mass $2M$ in gauge representation 
$R$ decaying into two massless gauge bosons in gauge representations
 $r$ and $r^\prime$~\cite{Beneke:2009rj,Becher:2003kh,Becher:2009qa,Becher:2009kw},
\begin{align}\label{eq:Gamma}
\Gamma(R) = \frac12 \gamma_{\rm cusp} \left[ \left( C_2(r) + C_2({r^\prime}) 
\right) 
\left( \log{4M^2\over \mu^2} - i\pi \right)
+ i\pi C_2(R)
\right] + \gamma^r + \gamma^{r^\prime} + \gamma^R  - 2{ \beta(g) \over g} \,.           
\end{align}
This makes the connection with the cusp anomalous dimension $\gamma_\mathrm{cusp}$ explicit.  Note that the coefficient of $\log 4M^2/ \mu^2$ is independent of the WIMP's spin and quantum numbers, demonstrating the universality of the Sudakov suppression for heavy WIMP annihilation.

The term in (\ref{eq:Gamma}) involving the beta function $\beta(g)=dg/d\log \mu$ appears due to the factor of $g^2$ in the operator definition.
Employing the expansion,
\begin{align}
\Omega &= \Omega_0{\alpha_2\over 4\pi} + 
\Omega_1\left(\alpha_2\over 4\pi\right)^2
+ \Omega_2\left(\alpha_2\over 4\pi\right)^3 + \dots \,,
\end{align} 
for the anomalous dimensions and beta function $\beta(\alpha_2)=d \alpha_2/d\log\mu$, we collect the 
coefficients necessary for resummation through NLL
order in Table~\ref{tab:functions}. 
From the one-loop results given in the first column of Table~\ref{tab:functions},
we recover \eqref{eq:Gisospin} from the ansatz in \eqref{eq:Gamma}, 
\emph{i.e.}, ${\hat \Gamma}^\prime = {\rm diag}(\Gamma(0), \Gamma(2) )$. 

\subsection{Sudakov Resummation}

\begin{table}[t]
\begin{center}
\small
\renewcommand{\arraystretch}{2}
\begin{tabular}[t]{c|c|c}
$\Omega$ & $\Omega_0$ & $\Omega_1$  \\
\hline
$\gamma_{\rm cusp}$  & $ 4$ & $\left(\frac{268}{9}-\frac{4}{3}\pi^2 \right)C_2(G)-\frac{80}{9}n_G-\frac{16}{9}$\\
\hline
$\gamma^R$ & $-2C_2(R) $ & -  \\
 \hline
 $\gamma^r, \gamma^{r\prime}$ & $- \left( {22 \over 3} - \frac16 -\frac43 n_G \right) $  & -  \\
 \hline
$-{\beta(\alpha_2) \over 2\alpha_2}$  & ${22 \over 3} - \frac16 -\frac43 n_G $ & $\frac{259}{6}-\frac{49}{3}n_G-\left(\frac{3}{10}-\frac{1}{5}\,
n_G\right)\frac{\alpha_1}{\alpha_2}-12\frac{\alpha_3}{\alpha_2}+\frac{3}{2}\frac{
\alpha_t}{\alpha_2}$  \\
\end{tabular}
\end{center}
\caption{\label{tab:functions}
Expansion coefficients of $\Omega = \sum_{n=0}^{\infty} \big( {\alpha_2 \over 4\pi}\big)^{n+1}  \Omega_n $ for the cusp and non-cusp anomalous dimensions and the $SU(2)_W$ beta function. 
The appearance of $\alpha_1 = g_1^2/4\pi$, $\alpha_3 = g_s^2/4\pi$ and $\alpha_t = Y_t^2/4\pi$ in $\beta_1$ (and higher order in $\gamma_{\rm cusp}$, $\gamma^R$ and $\gamma^r,\gamma^{r\prime}$) 
complicates the analysis beyond LL order. 
}
\end{table}

Let us consider the solution for coefficient scale evolution 
governed by (\ref{eq:RGE1}). We write
\begin{align}\label{eq:RGsolution}
{\vec c}(\mu_L)  = 
{\hat S}(\mu_L,\mu_H) \, {\vec c}(\mu_H)= S_{\rm cusp}(\mu_L,\mu_H)\, {\hat S}_R(\mu_L, \mu_H)\, {\vec c}(\mu_H)\,,
\end{align}
where the function $S_{\rm cusp}$ accounts for the universal scale evolution from the cusp anomalous dimension, while the matrix 
$\hat S_R$ accounts for scale evolution from the isospin-dependent non-cusp anomalous dimension.
To LL accuracy, the solution reads
\begin{align}\label{eq:SLL}
{S}_{\rm cusp}^{\rm LL} = \exp \Big[{4\pi \over \alpha_2(\mu_L) }   { \gamma_{{\rm cusp},0} \over \beta_0^2 } \Big\{ r - 1 - r \log r \Big\}  + 
{ 2 \gamma_{{\rm cusp},0} \over \beta_0} \log {\mu_H \over 2M} \log r \Big] \, , \quad S_0^{\rm LL} = S_2^{\rm LL} = 1 \,, 
\end{align}
where $r = \alpha_2(\mu_L) / \alpha_2(\mu_H)$ and $S_0$, $S_2$ are the diagonal elements of $\hat S_R$ in the isospin basis. In the  
(non-isospin) basis of operators $O_{1,2}$, we have
\begin{empheq}[box={\mybluebox[5pt]}]{align}
\label{eq:Sgen}
{\hat S} = S_{\rm cusp}  \left(\begin{array}{cc} 
S_0 & \ \frac23 (S_0 - S_2) \\
0 & \ S_2
\end{array}\right) \, ,
\end{empheq}
such that mixing effects enter only at NLL order. 

Let us make the explicit connection between the cusp anomalous dimension and the
Sudakov double log appearing in the charged WIMP annihilation
amplitude ${\cal M}^{+- \to \gamma \gamma}$ in
\eqref{eq:onestep1PICharged}. 
Writing $r$ as a
series in $\alpha_2(\mu_L)$ 
we find 
\begin{align}\label{eq:SLLexp}
{S}_{\rm cusp}^{\rm LL}(m_W,2M) =1 - {\alpha_2 \over 4\pi} 2 \gamma_{{\rm cusp},0} \log^2 {m_W \over 2M} + \left( {\alpha_2 \over 4\pi} \right)^2 
2 \gamma_{{\rm cusp},0}^2 \log^4 {m_W \over 2M} + \dots + \order(\alpha_2^3) \, ,
\end{align}
where in this expression $\alpha_2 = \alpha_2(m_W)$, and the ellipsis
denotes non-leading log pieces omitted above. Comparing with ${\cal M}^{+- \to \gamma \gamma}/2e^2$ in \eqref{eq:onestep1PICharged}, we see that
the Sudakov double log is exactly recovered with its coefficient tied
to the cusp anomalous dimension as expected. 

The full NLL solution can be straightforwardly derived using the coefficients given in Table~\ref{tab:functions}.
Note that beyond LL order, the running of couplings $\alpha_1$, $\alpha_s$
and $\alpha_t$ enter the RGE through $\beta_1$ which, however, appears
only at $\order(\alpha_2^3)$ in $\hat{S}$ (see,
\emph{e.g.} Ref.~\cite{Almeida:2014uva}). The smallness of $\alpha_2$ thus
implies that to good approximation we may investigate the numerical
impact of NLL resummation with these couplings kept
constant.\footnote{We verify numerically that varying the fixed values
of $\alpha_1$, $\alpha_s$ and $\alpha_t$ within appropriate ranges has
a negligible effect on $\hat{S}$.} The impact of LL and NLL resummation is investigated below (see Fig.~\ref{fig:wpmBands}).

For the present study, we focus on LL accuracy, employing the solution for ${\hat S}$ specified by \eqref{eq:SLL}
and the one-loop hard scale coefficients $c_i(\mu_H)$ given in
\eqref{eq:renormc}. Our numerical investigation of corrections at LL and NLL orders indicate good perturbative convergence. The framework presented here can be readily employed for a detailed investigation of higher-order resummation relevant for WIMPs with mass in the multi-TeV range and beyond.

\section{Weak Scale Matching \label{sec:lowmatch}}

Having solved the RGEs in electroweak symmetric SCET, we now have the Wilson coefficients of the annihilation operators at the low scale $\mu_L \sim m_W$ in terms of those at the high scale $\mu_H\sim 2M$.  The final step is to match operators in this EFT, expressed in the field basis of broken electroweak symmetry, onto the quantum mechanical Hamiltonian discussed in (\ref{eq:H}) above. This matching will determine the elements of the RG improved $W$ matrix, which is convolved with the Sommerfeld matrix to obtain the annihilation cross section. The first task is to derive the Feynman rules for electroweak broken SCET that will then be used to compute one-loop corrections for the SCET side of the matching condition.

\subsection{Electroweak Broken SCET Feynman Rules}
These Feynman rules are the exact analog of what was discussed in Sec.~\ref{sec:EWSymSCETFeynRules} except we are now working in the electroweak broken phase.  For simplicity, we again specialize to the isospin $j=1$ case, $(t^a)^{bc} = if^{bac}$. The operators are defined as in (\ref{eq:originalbasis}), but with gauge fields written in terms of $\gamma$, $Z^0,$ and $W^\pm$, introducing a dependence on $s_W^2 \equiv \sin^2 \theta_W$.  Note that we have followed the same convention as above, defining the $c_i$ Wilson coefficients to be dimensionless (a $1/M$ factor appears in the Lagrangian \eqref{eq:Leff}).  

The Feynman rules for two gauge boson emission are 
\\
\begin{align}
  \parbox{40mm}{
    \begin{fmfgraph*}(100,80)
      \fmfleftn{l}{2}
      \fmfrightn{r}{2}
      \fmf{double}{l2,a}
      \fmf{double}{l1,a}
      \fmf{photon}{a,r2}
      \fmf{photon}{a,r1}
      \fmffreeze
      \fmflabel{$p,\mu$}{r2}
      \fmflabel{$p^\prime,\nu$}{r1} 
      \fmflabel{$0$}{l2}
      \fmflabel{$0$}{l1}
      \fmfv{decor.shape=circle,decor.filled=empty,decor.size=.1w}{a}
    \end{fmfgraph*}
  }
& = g_\perp^{\mu\nu} ( 2i s_W^2 c_1 ) \, ,
\nonumber
\\[15mm] 
  \parbox{40mm}{
    \begin{fmfgraph*}(100,80)
      \fmfleftn{l}{2}
      \fmfrightn{r}{2}
      \fmf{double}{l2,a}
      \fmf{double}{l1,a}
      \fmf{zigzag}{a,r2}
      \fmf{zigzag}{a,r1}
      \fmffreeze
      \fmflabel{$p,\mu, +$}{r2}
      \fmflabel{$p^\prime,\nu, -$}{r1} 
      \fmflabel{$0$}{l2}
      \fmflabel{$0$}{l1}
      \fmfv{decor.shape=circle,decor.filled=empty,decor.size=.1w}{a}
    \end{fmfgraph*}
  }
& = \quad 
  \parbox{40mm}{
    \begin{fmfgraph*}(100,80)
      \fmfleftn{l}{2}
      \fmfrightn{r}{2}
      \fmf{double}{l2,a}
      \fmf{double}{l1,a}
      \fmf{zigzag}{a,r2}
      \fmf{zigzag}{a,r1}
      \fmffreeze
      \fmflabel{$p,\mu, -$}{r2}
      \fmflabel{$p^\prime,\nu, +$}{r1} 
      \fmflabel{$0$}{l2}
      \fmflabel{$0$}{l1}
      \fmfv{decor.shape=circle,decor.filled=empty,decor.size=.1w}{a}
    \end{fmfgraph*}
  } 
= g_\perp^{\mu\nu} [ 2i ( c_1 +c_2 )] \, ,
\nonumber
\\[15mm] 
  \parbox{40mm}{
    \begin{fmfgraph*}(100,80)
      \fmfleftn{l}{2}
      \fmfrightn{r}{2}
      \fmf{double}{l2,a}
      \fmf{double}{l1,a}
      \fmf{photon}{a,r2}
      \fmf{photon}{a,r1}
      \fmffreeze
      \fmflabel{$p,\mu$}{r2}
      \fmflabel{$p^\prime,\nu$}{r1} 
      \fmflabel{$+$}{l2}
      \fmflabel{$-$}{l1}
      \fmfv{decor.shape=circle,decor.filled=empty,decor.size=.1w}{a}
    \end{fmfgraph*}
  }
& = g_\perp^{\mu\nu} [2i s_W^2 (c_1+c_2)] \, ,
\nonumber
\\[15mm]
 \parbox{40mm}{
    \begin{fmfgraph*}(100,80)
      \fmfleftn{l}{2}
      \fmfrightn{r}{2}
      \fmf{double}{l2,a}
      \fmf{double}{l1,a}
      \fmf{zigzag}{a,r2}
      \fmf{photon}{a,r1}
      \fmffreeze
      \fmflabel{$p,\mu,\pm$}{r2}
      \fmflabel{$p^\prime,\nu$}{r1} 
      \fmflabel{$0$}{l2}
      \fmflabel{$\pm$}{l1}
      \fmfv{decor.shape=circle,decor.filled=empty,decor.size=.1w}{a}
    \end{fmfgraph*}
  }
& = \quad
 \parbox{40mm}{
    \begin{fmfgraph*}(100,80)
      \fmfleftn{l}{2}
      \fmfrightn{r}{2}
      \fmf{double}{l2,a}
      \fmf{double}{l1,a}
      \fmf{photon}{a,r2}
      \fmf{zigzag}{a,r1}
      \fmffreeze
      \fmflabel{$p,\mu$}{r2}
      \fmflabel{$p^\prime,\nu,\pm$}{r1} 
      \fmflabel{$0$}{l2}
      \fmflabel{$\pm$}{l1}
      \fmfv{decor.shape=circle,decor.filled=empty,decor.size=.1w}{a}
    \end{fmfgraph*}
  }  
= g_\perp^{\mu\nu} (-i s_W c_2 ) \,, 
\end{align}
\\
where we draw double straight lines for the heavy WIMP initial states (now being careful to distinguish the electric charge), 
wavy lines for the photon, and jagged lines for the $W^\pm$ gauge bosons.  
For an additional massive hardcollinear emission from the $O_i$ vertex,
\\
\begin{align}
 \parbox{40mm}{
    \begin{fmfgraph*}(100,80)
      \fmfleftn{l}{2}
      \fmfrightn{r}{4}
      \fmf{double}{l2,a}
      \fmf{double}{l1,a}
      \fmf{zigzag}{a,r4}
      \fmf{photon}{a,r1}
      \fmffreeze
      \fmf{zigzag}{a,r3}
      \fmflabel{$k,\mu,+$}{r4}
      \fmflabel{$q,\rho,-$}{r3}
      \fmflabel{$p^\prime,\nu$}{r1} 
      \fmflabel{$0$}{l2}
      \fmflabel{$0$}{l1}
      \fmfv{decor.shape=circle,decor.filled=empty,decor.size=.1w}{a}
    \end{fmfgraph*}
  }
&= \quad ig (2c_1) \left( {\bar{n}^\mu \over \bar{n}\cdot k} 
g_{\perp}^{\rho\nu} 
- {\bar{n}^\rho \over \bar{n}\cdot q}g_{\perp}^{\nu\mu} 
\right) \,,
\end{align} 
\\
with a similar rule for two anti-hardcollinear emissions with $n\leftrightarrow 
\bar{n}$ as before.  
The interaction of soft gauge fields with heavy scalars is again given by the usual 
result, 
\\
\begin{align}
  \parbox{35mm}{
    \begin{fmfgraph*}(100,60)
      \fmfleftn{l}{3}
      \fmfrightn{r}{3}
      \fmftopn{t}{3}
      \fmf{double}{l1,a}
      \fmf{double}{r1,a}
      \fmffreeze
      \fmf{photon}{t2,a}
      \fmflabel{$\pm$}{l1}
      \fmflabel{$\pm$}{r1}
      \fmfv{label=$\mu$,label.angle=90}{t2}
    \end{fmfgraph*}
  }
&= \pm ie \,,
\quad 
 \parbox{35mm}{
    \begin{fmfgraph*}(100,60)
      \fmfleftn{l}{3}
      \fmfrightn{r}{3}
      \fmftopn{t}{3}
      \fmf{double}{l1,a}
      \fmf{double}{r1,a}
      \fmffreeze
      \fmf{zigzag}{t2,a}
      \fmflabel{$0$}{l1}
      \fmflabel{$\pm$}{r1}
      \fmfv{label=$\mu$,label.angle=90}{t2}
    \end{fmfgraph*}
  }
= \mp i g \,,
\quad 
 \parbox{35mm}{
    \begin{fmfgraph*}(100,60)
      \fmfleftn{l}{3}
      \fmfrightn{r}{3}
      \fmftopn{t}{3}
      \fmf{double}{l1,a}
      \fmf{double}{r1,a}
      \fmffreeze
      \fmf{zigzag}{t2,a}
      \fmflabel{$\pm$}{l1}
      \fmflabel{$0$}{r1}
      \fmfv{label=$\mu$,label.angle=90}{t2}
    \end{fmfgraph*}
  }
= \mp i g \,.
\end{align} 
\\
Note that rules involving the $Z^0$ can be inferred by changing a photon to a $Z^0$ and multiplying the coupling by $c_W/s_W$.   

Armed with these Feynman rules, we may now compute the full
one-loop matrix element for neutral and charged heavy WIMP
annihilation to photons.  
As in Sec.~\ref{sec:hardmatching}, matching must be performed between physical amplitudes, requiring onshell wavefunction
factors for the external states. For the gauge field, these are the
same as in the full theory, and the combination needed for this
calculation $(Z_1^W)^2(Z_2^W)^{-2}$ is given in (\ref{eq:ZW}) above.
For the heavy neutral field, 
\be
 Z_{\phi_0} = 1 + {g^2\over (4\pi)^2} \bigg[ 
{4\over \epsilon} - 8 \log{m_W\over \mu} 
\bigg]
 \,, 
\ee
while for the heavy charged field, 
\be
 Z_{\phi_\pm} = 1 + 
{g^2\over (4\pi)^2}\bigg[ {4\over \epsilon}  - 4\log{m_W\over\mu} 
- 4s_W^2\log{m_\gamma\over \mu} 
- 4 c_W^2 \log{m_Z \over \mu} \bigg]  
\,. 
\ee
Note that since electroweak symmetry is broken, the charged and neutral states are split
due to one-loop corrections from the gauge bosons; (\ref{eq:MassSplitting}) also applies in the EFT.  

\subsection{WIMP Annihilation in Electroweak Broken SCET} 
\begin{figure}[h!]
\begin{center} 
\input{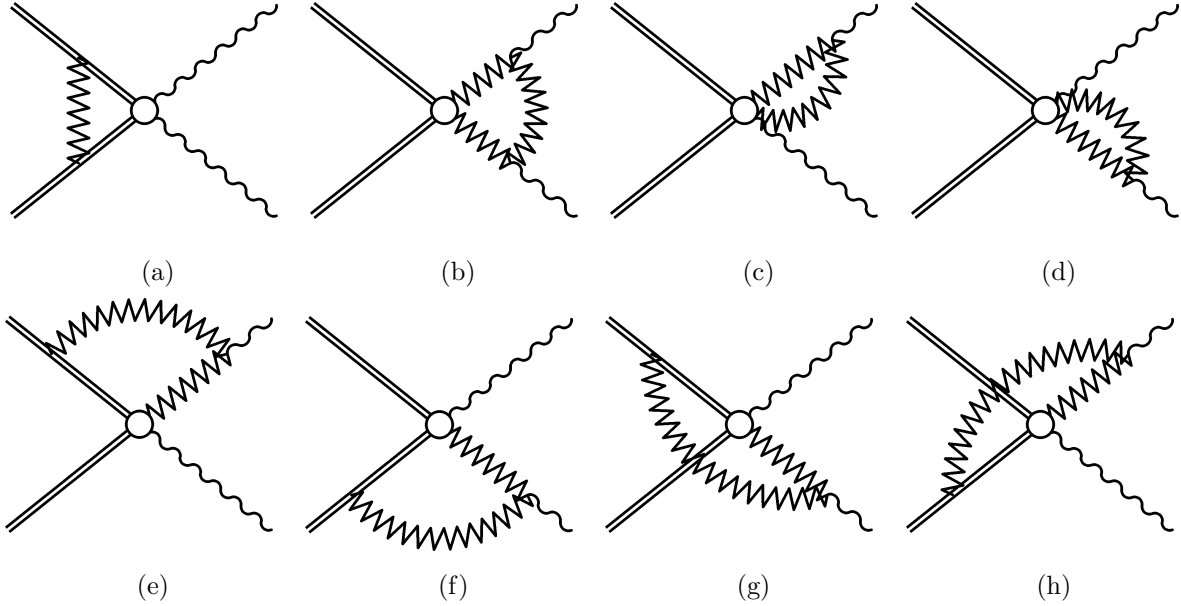}
\caption{One-loop contributions to matrix elements of $O_i$ for neutral WIMPs.  Double straight lines are heavy WIMPs, wavy lines are photons, and jagged lines are $W^\pm$ bosons.}
\label{fig:neutralEWSB}
\end{center}
\end{figure}

All that remains to obtain the desired result are the finite terms from matching at one loop in electroweak broken
SCET.  We begin by providing results for neutral WIMP annihilation.  The diagrams are 
given in Fig.~\ref{fig:neutralEWSB}.  Using the Feynman rules of the previous section 
we proceed to compute the one-loop matrix element for annihilation of two neutral heavy particles into two photons.  
Including the appropriate onshell renormalization constants, we find
\begin{align} \label{eq:neutralEW}
& i{\cal M}^{00\to\gamma\gamma}
= 2 i e^2 c_1(\mu) 
+ {i e^2 g^2 \over (4\pi)^2} \bigg\{ 
c_1(\mu) \bigg[ 
C_\text{potential}
 - 16 \log^2{m_W\over \mu} 
+ 32 \log{2M\over \mu} \log{m_W\over \mu} \nl
&\quad 
 - 16 i \pi \log{m_W\over \mu} - {4\pi^2\over 3} \bigg] 
+ c_2(\mu) \bigg[ 
C_\text{potential}
+ 16(1-i\pi) \log{m_W\over \mu} \bigg]
\bigg\} \,,
\end{align}
where the only dependence on the threshold is captured by $C_\text{potential}$, 
which is given in (\ref{eq:Cpotential}). 

\begin{figure}[h!]
\begin{center} 
\input{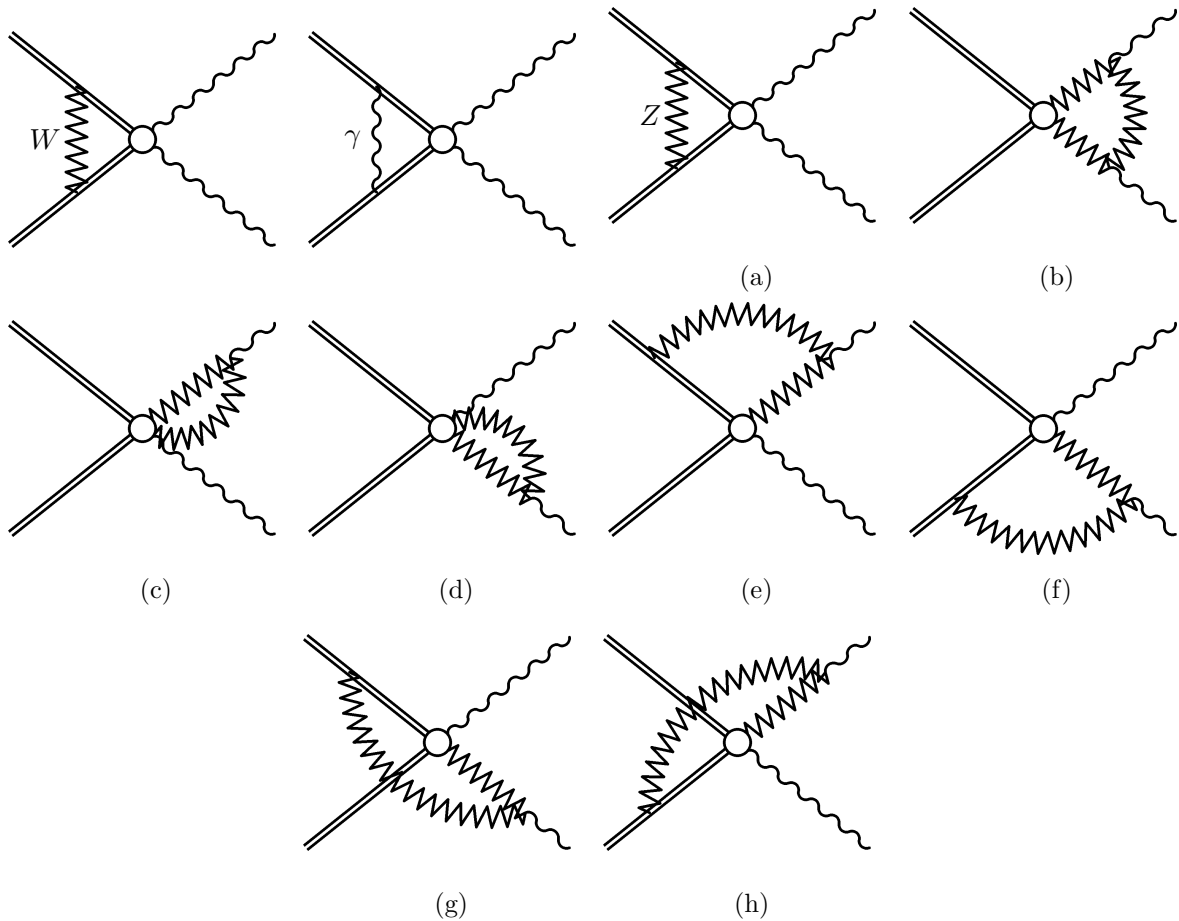}
\caption{One-loop contributions to matrix elements of $O_i$, for charged WIMP 
annihilation.  The wavy lines are photons, and the jagged lines are $W^\pm$, except when explicitly labeled as a $Z^0$.
\label{fig:chargedEWSB}}
\end{center}
\end{figure}

Let us compute the diagrams in Fig.~\ref{fig:chargedEWSB} relevant to the 
charged annihilation at $k^\mu = \delta v^\mu$, 
\emph{i.e.}, the threshold annihilation for charged states (in comparison to above, we 
include factors of $g$ and $s_W$).  The renormalized amplitude is 
\begin{align}\label{eq:chargedEW}
&i{\cal M}^{+-\to\gamma\gamma}
= 2 i e^2 [c_1(\mu) +c_2(\mu)]
+ {i e^2 g^2 \over (4\pi)^2} \bigg\{ 
c_1(\mu) {8\pi M \over m_W + \sqrt{-2M\delta-i0} } + [c_1(\mu)+c_2(\mu)] {8\pi s_W^2 M\over m_\gamma} 
\nl
&\quad 
+ [c_1(\mu)+c_2(\mu)] {8\pi c_W^2 M\over m_Z}
+ c_1(\mu) \bigg[ -{4\pi^2\over 3} + 32\log{2M\over \mu} \log{m_W\over \mu} 
- 16 i \pi \log{m_W\over \mu} 
- 16\log^2{m_W\over \mu} 
\bigg]
\nl
&\quad
+  c_2(\mu) \bigg[ -{4\pi^2\over 3} 
+ 32\log{2M\over \mu} \log{m_W\over \mu} 
- 8 i \pi \log{m_W\over \mu} 
- 16\log^2{m_W\over \mu} - 8\log{m_W\over \mu} 
\bigg]
\bigg\} \,.
\end{align}
Note that we have taken $n_G=3$ in both (\ref{eq:neutralEW}) and (\ref{eq:chargedEW}).

\subsection{Collinear Anomaly}
\label{sec:CollinearAnomaly}
In evaluating the amplitudes, \emph{e.g.}, diagram (c) in Figs.~\ref{fig:neutralEWSB} 
and \ref{fig:chargedEWSB}, care must be taken to subtract a nonvanishing 
soft region contribution from the collinear momentum integral.  This nontrivial
subtraction is a remnant of 
nonfactorization between the collinear sectors~\cite{Becher:2003qh}, 
and manifests itself as residual 
dependence of the low-energy matrix elements (\ref{eq:neutralEW}), (\ref{eq:chargedEW}) 
on $\log{M/\mu}$, appearing at leading power in $m_W/M$. 
For problems involving a single IR scale, 
this residual dependence can be factorized to all orders in perturbation 
theory~\cite{Becher:2010tm, Becher:2011xn, Chiu:2011qc, Becher:2011pf, Chiu:2012ir, Becher:2012qa}.
In the present case, we take 
\begin{align}
c_i(\mu) \to c_i(\mu) \left( 4M^2 \over \mu^2 \right)^{-\frac12 F(m_W,\mu)} \,, 
\end{align}
where at leading order, 
\be
F(m_W,\mu) = {\alpha_2\over 4\pi} {F_0} \log{\mu^2\over m_W^2} \,, \quad
F_0 = 4 C_2(j) \,.
\ee
The interplay of this so-called collinear anomaly and electroweak symmetry breaking 
will modify this structure beyond one loop.  This order of precision is beyond
phenomenological importance in the present application, and a more detailed
exposition is left to future work.  

\subsection{Weak Scale Matching Results} 
The quantum mechanical side of the matching computation is identical to that obtained above
in Sec.~\ref{sec:WQM}, and can be used to compute the analog of (\ref{eq:onestepw}), which was derived by directly matching with the full electroweak broken theory.
The absorptive part of the potential, including the effects of resummation, are thus 
\begin{empheq}[box={\mybluebox[5pt][5pt]}]{align}
\label{eq:twostepc}
w_{\pm} &= -{4\pi\alpha^2\over M^2} 
\left( 4M^2 \over \mu^2 \right)^{-F(m_W,\mu)}
\bigg\{ |c_1+c_2|^2 
+ {\alpha_2 \over 4\pi }\bigg[ |c_1+c_2|^2 \bigg( -16\log^2{m_W\over \mu}
\nl
&\quad
 - 8\log{m_W\over \mu}  -{4\pi^2\over 3} \bigg) 
+ {\rm Re} \left( (c_1+c_2)^* c_1 ( 8 -8i\pi )\log{m_W\over \mu} \right) \bigg]
+ \order(\alpha^2) 
\bigg\} 
\,,
\nl
w_{\pm;0} &= -{4\pi\alpha^2\over M^2}
\left( 4M^2 \over \mu^2 \right)^{-F(m_W,\mu)}
 \bigg\{ c_1 (c_1+c_2)^* 
+ {\alpha_2\over 4\pi}\bigg[ 
c_1 (c_1+c_2)^*\bigg( -16\log^2{m_W\over \mu} 
\nl
&\quad
 - 12\log{m_W\over\mu} +4i\pi\log{m_W\over\mu} 
- {4\pi^2\over 3} \bigg)
+ |c_1|^2\bigg( (4+4i\pi)\log{m_W\over \mu} \bigg)
\nl
&\quad
+ |c_1+c_2|^2\bigg(  (8-8i\pi) \log{m_W\over \mu} \bigg) 
\bigg]
+ \order(\alpha^2) 
\bigg\}
\,,
\nl
w_{0} &= -{4\pi\alpha^2\over M^2}
\left( 4M^2 \over \mu^2 \right)^{- F(m_W,\mu)}
\bigg\{ |c_1|^2 
+ {\alpha_2\over 4\pi}\bigg[ 
{\rm Re}\bigg( (c_1+c_2)^*c_1 (16+16i\pi) \log{m_W\over \mu} \bigg)
\nl
&\quad 
+ |c_1|^2 \bigg( -16\log^2{m_W\over\mu} -16\log{m_W\over \mu} 
-{4\pi^2\over 3}
\bigg) 
\bigg] + \order(\alpha^2) 
\bigg\}
\,,
\end{empheq}
where $c_i(\mu)$ are the solutions 
(\ref{eq:RGsolution}) to the RG evolution 
equation, with high scale coefficients (\ref{eq:renormc}), 
and $w_{ij}$ are defined in (\ref{eq:Wtree}) 
using plane-wave external states.  
These expressions accomplish a complete factorization of the 
scales $2M$ and $m_W$, and systematically resum the large logarithms 
of perturbation theory.

\section{Implications \label{sec:pheno}}

Having completed the high scale matching (\ref{eq:renormc}), RG running (\ref{eq:Sgen}) and finally low scale matching (\ref{eq:twostepc}), we may proceed to use the Hamiltonian to compute interesting physical observables and investigate the impact of perturbative corrections.

\begin{figure}[h!]
\begin{centering}
\includegraphics[width=.65 \textwidth]{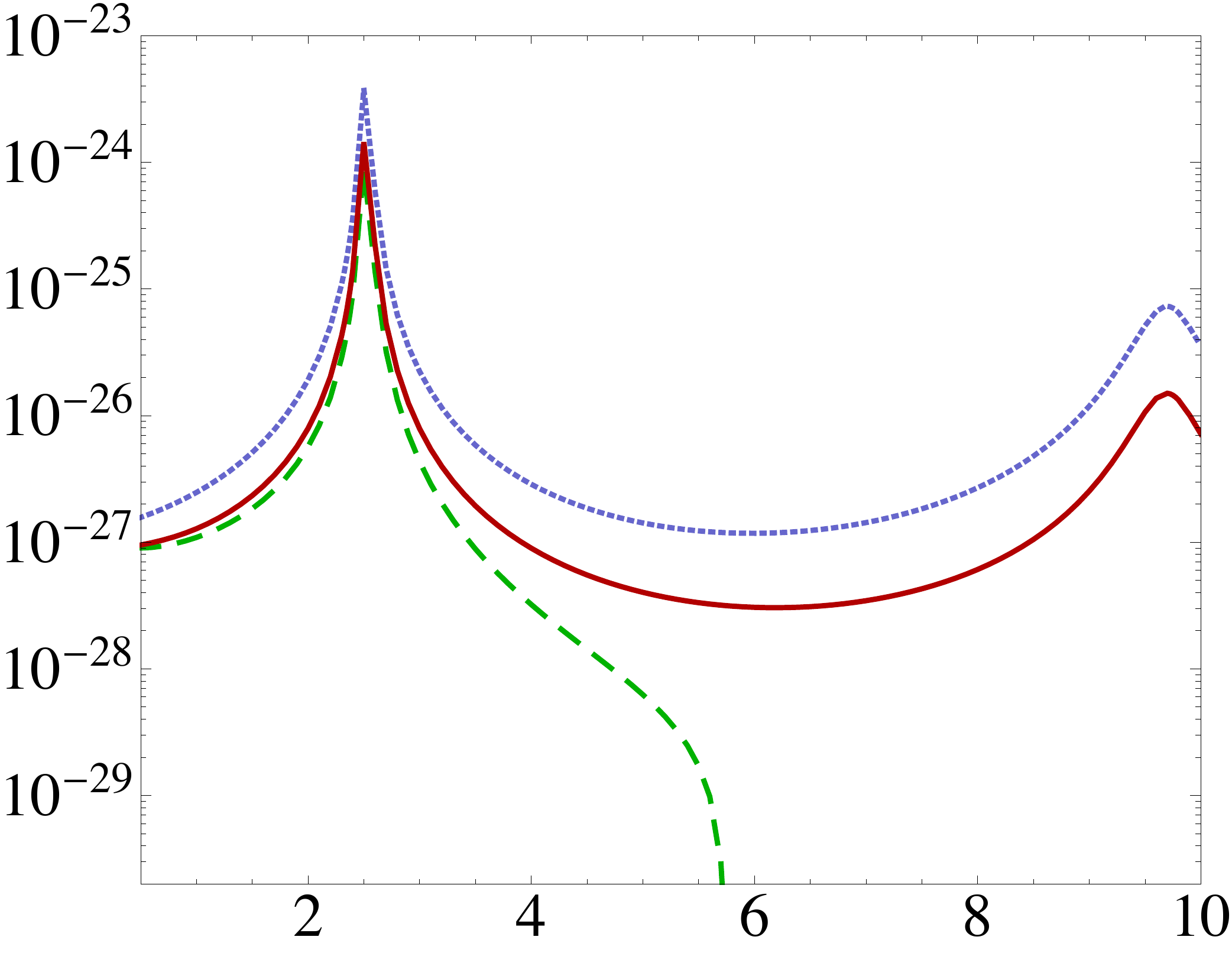}
\hspace*{10mm}
\begin{minipage}{0cm}
\vspace*{1.6cm}\hspace*{-7.1cm}{\Large $M\,\,\big[\text{TeV} \big]$}
\vspace*{0.5cm}
\end{minipage}
\begin{minipage}{0cm}
\vspace*{-8.5cm}\hspace*{-13.2cm}\rotatebox{90}{{\Large $\sigma\,v\,\,\big[\text{cm}^3/\text{s}\big]$}}
\end{minipage}
\vspace{-12pt}
\caption{Sommerfeld enhanced WIMP annihilation cross sections for $\phi\,\phi \rightarrow \gamma\,\gamma$ employing three approximations.  
The fixed $\order(\alpha^2)$ result is shown in dotted blue. The fixed $ \order(\alpha^3)$ result, including the first non-vanishing $\order(\alpha^4)$ contribution to $w_{00}$, is shown in dashed green. The LL resummed result, including one-loop matching coefficients
at the hard and weak scales and resummation of the collinear 
anomaly contribution, is shown in solid red.}
\label{fig:treeVsNLOVsResummed}
\end{centering}
\end{figure}

Figure \ref{fig:treeVsNLOVsResummed} shows the Sommerfeld
enhanced annihilation cross section to line photons  for three
approximations, taking $\delta = 0.17 \, \text{GeV}$ and $v = 10^{-3}$ as above. The blue dotted and green dashed lines are fixed order results at $\order(\alpha^2)$ and $\order(\alpha^3)$, respectively, with the latter also including the first non-vanishing $\order(\alpha^4)$ contribution to $w_{00}$. The red solid line is the result including LL resummation, one-loop matching coefficients
at the high and weak scales, and resummation of the collinear 
anomaly contribution. The uncertainty from scale variation would not be resolved on this log plot, hence we only show the central value and discuss perturbative uncertainties below.
As previously discussed the fixed $\order(\alpha^3)$ result (green dashed) becomes negative for $M \gtrsim 6 \, {\rm TeV}$, indicating a breakdown in perturbation theory.

\begin{figure}[h!]
\begin{centering}
\includegraphics[width=.62 \textwidth]{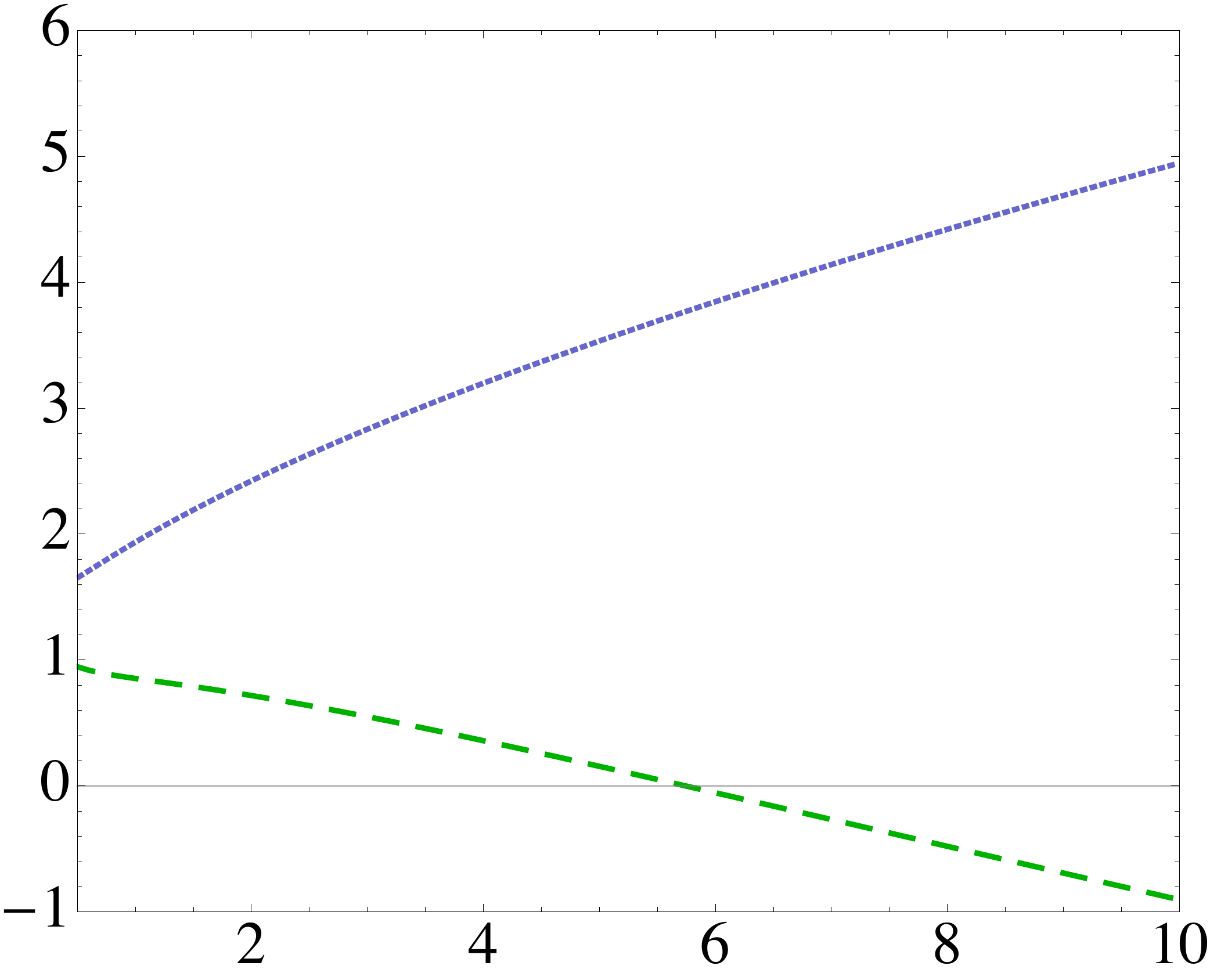}
\hspace*{10mm}
\begin{minipage}{0cm}
\vspace*{1.6cm}\hspace*{-7.1cm}{\Large $M\,\,\big[\text{TeV} \big]$}
\vspace*{0.5cm}
\end{minipage}
\begin{minipage}{0cm}
\vspace*{-13.cm}\hspace*{-8.2cm}{\Large $(\sigma\,v)_\text{tree}/(\sigma\,v)_\text{LL}$}
\end{minipage}
\begin{minipage}{0cm}
\vspace*{-5.3cm}\hspace*{-8.2cm}{\Large $(\sigma\,v)_\text{1-loop}/(\sigma\,v)_\text{LL}$}
\end{minipage}
\vspace{-12pt}
\caption{This plot shows the ratio of $(\sigma\,v)_\text{tree}/(\sigma v)_\text{LL}$ (blue dotted) and $(\sigma\,v)_\text{1-loop}/(\sigma v)_\text{LL}$ (green dashed) including the effects of the Sommerfeld enhancement.}
\label{fig:resummedVs1StepPercentDiff}
\end{centering}
\end{figure}

There is a robust suppression of the resummed result due to the LL correction from the (universal) cusp anomalous dimension. We give the ratios of the Sommerfeld enhanced fixed order cross sections to the resummed cross section, $ (\sigma v)_\text{tree} / (\sigma v)_\text{LL}$ and $(\sigma v)_\text{1-loop}/(\sigma v)_\text{LL}$, in Fig.~\ref{fig:resummedVs1StepPercentDiff}. At $M= 3\, {\rm TeV}$ the resummed result is suppressed by a factor of $\sim 3$ with respect to tree level.

To illustrate the impact of higher order perturbative corrections, let
us investigate the residual renormalization scale dependence of the
absorptive part of the potential at LL and NLL accuracy. We focus here
on $w_\pm$ which has the largest impact on the neutral WIMP annihilation cross section to photons.
For LL order, we include the LL solution to the RG evolution and tree-level matching coefficients at the hard and intermediate scales, but neglect the collinear anomaly contribution. 
For NLL order, we include the NLL solution to the RG evolution, tree-level matching coefficients at the hard and intermediate scales and full resummation of the collinear anomaly contribution.

\begin{figure}[h!]
\begin{centering}
\includegraphics[width=.65 \textwidth]{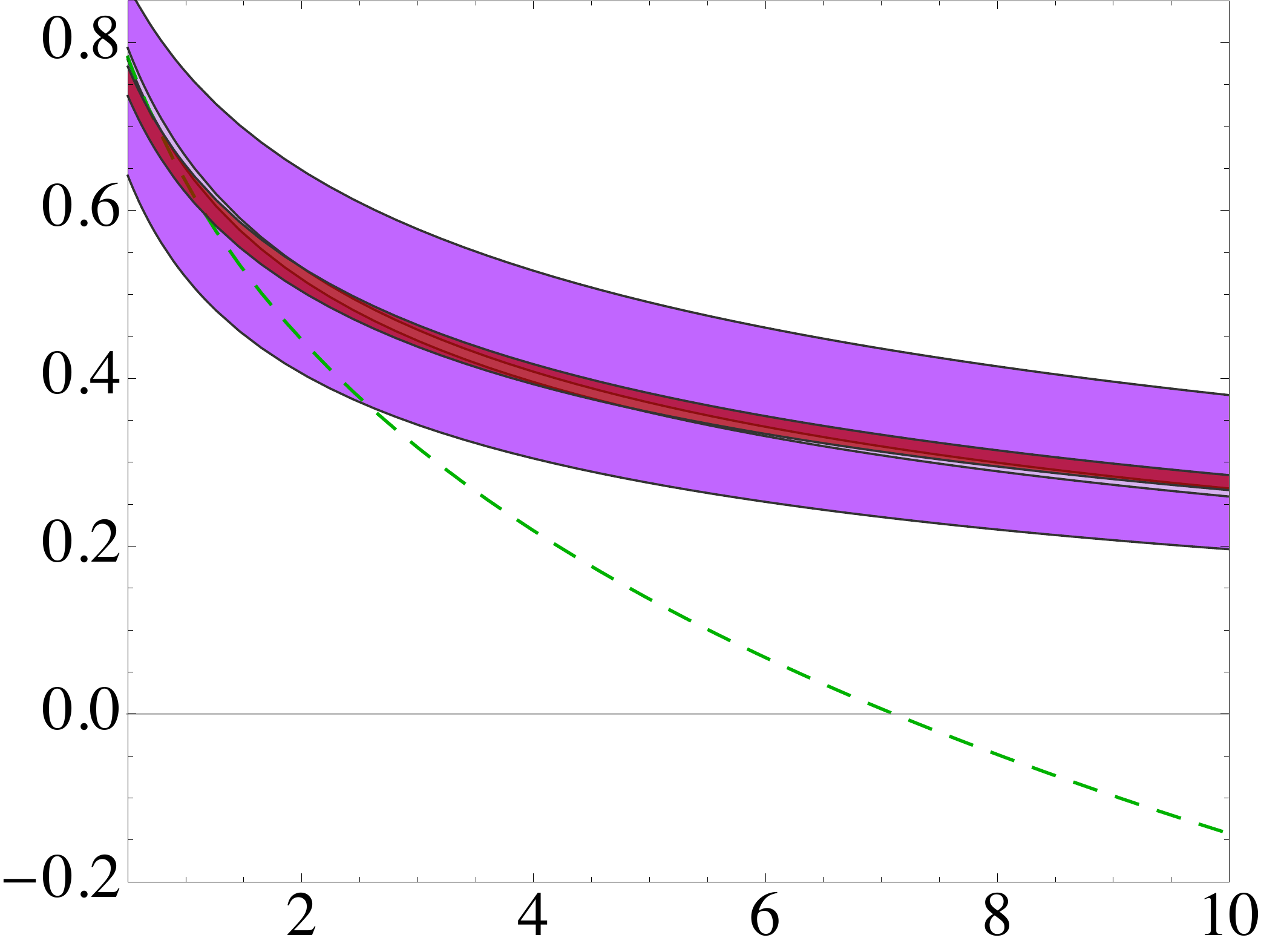}
\hspace*{10mm}
\begin{minipage}{0cm}
\vspace*{1.6cm}\hspace*{-7.4cm}{\Large $M\,\,\big[\text{TeV} \big]$}
\vspace*{0.5cm}
\end{minipage}
\begin{minipage}{0cm}
\vspace*{-8.5cm}\hspace*{-13.0cm}\rotatebox{90}{{\Large $-\frac{M^2}{\pi\,\alpha^2}\,w_\pm$}}
\end{minipage}
\vspace{-12pt}
\caption{The LL (purple) and NLL (gray) results for $-{M^2  } \,w_\pm/ \pi \alpha^2$ with estimated error bands combined from varying $m_W< \mu_L < 2 m_W$ and $M < \mu_H < 4 M$. For comparison, we also include the fixed $\order(\alpha^3)$ result (dashed green line), and the LL resummed result (red band) employed for $\sigma v$ in Fig.~\ref{fig:treeVsNLOVsResummed} above.}
\label{fig:wpmBands}
\end{centering}
\end{figure}

The results of this study are shown in Fig.~\ref{fig:wpmBands} where we plot $w_\pm$ in units of ${M^2  } / \pi \alpha^2$ so that the tree-level result is unity. The purple and grey bands are the LL and NLL results, respectively, where the uncertainty is from the combined variation of scales $m_W/2 < \mu_L < 2 m_W$ and  $M < \mu_H < 4 M$. For comparison, we also include the fixed $\order(\alpha^3)$ result (dashed green line), and the LL resummed result (red band) employed for $\sigma v$ in Fig.~\ref{fig:resummedVs1StepPercentDiff} above. The fixed order result has no explicit $\mu$ dependence, while the uncertainty for the red band is from the combined variation of scales $m_W/2 < \mu_L < 2 m_W$ and  $M < \mu_H < 4 M$.
The sizable uncertainty in the LL result (purple band) is due to the scale variation of the Sudakov double log, which cancels at NLL order with the variation of the $\order(\alpha)$ contribution from the collinear anomaly.

The resummed results capture the large $\alpha \log^2 {2M \over m_W}$ contribution through scale evolution of the hard matching coefficients $c_i(\mu)$, which enter quadratically in (\ref{eq:twostepc}). The fixed order result, on the other hand, has the large $\alpha \log^2 {2M \over m_W}$ contribution but appearing only linearly in $w_\pm$. For $M \gtrsim 7 \, {\rm TeV}$ the missing contributions result in $w_\pm$ becoming positive ($-{M^2  } \,w_\pm/ \pi \alpha^2$ becoming negative) which translates to a negative $\sigma v$ in Fig.~\ref{fig:resummedVs1StepPercentDiff} above. The resummation of large logarithms is necessary for control of perturbative corrections. 

\section{Summary\label{sec:summary}}

We have constructed a general EFT framework to
analyze heavy WIMP annihilation.  The factorization accomplished in
(\ref{eq:twostepc}) provides a systematically improvable  framework in
which to compute annihilation observables.   By separating the WIMP,
$M$, and electroweak,  $m_W$, scales, the EFT allows hard scale
matching conditions to be efficiently  computed in the electroweak
symmetric theory, while low-scale matching conditions and
long-distance wavefunction analysis may be performed in simpler
effective theories.  

At the same time, large logarithms that would otherwise lead to a
breakdown in perturbation  theory are systematically resummed by
solving the RGEs derived from the effective theory operators  in the
intermediate, soft collinear, effective theory.  In particular, a
universal  suppression of heavy WIMP annihilations is traced to the
cusp anomalous dimension governing effective theory operators.   We
provided details of the operator construction, hard scale matching,
and renormalization  of this effective theory. 

Below the electroweak scale, we mapped the problem to the relevant
quantum mechanical  Hamiltonian describing the nonrelativistic WIMP
system.  The relevant matching conditions in this effective theory
were computed, and used to derive expressions for the  absorptive part
of the potential representing the chosen annihilation channel.   This
two-step matching procedure recovers the results of a one-step
matching procedure at  fixed order in perturbation theory, but
systematically resums large logarithms.   Having fully determined the
low energy theory in a controlled perturbative expansion,  we computed
an illustrative observable represented by the low-velocity
annihilation rate to  two photon final states.   

The EFT framework presented here can be  applied to
a broad class of models and signatures.   Details of the particular
UV completion are encoded in the hard scale matching coefficients,
while heavy WIMP spin symmetry implies the existence of general features that are associated
with the remaining physical scales.   In particular, the dominant effect from including the loop corrections derives from a universal factor that is independent of the spin and electroweak quantum numbers of the WIMP.   Disentangling the different
energy scales in a sequence of effective theories allows the separate
treatment of physical effects  associated with the hard annihilation
process, the Sudakov suppression, and  the Sommerfeld enhancement of
annihilation observables.   Subleading perturbative, power and
velocity corrections may be systematically incorporated.  

As a concrete application, we focused attention on a heavy scalar
$SU(2)_W$ triplet annihilating into photons.  While fixed order
perturbation theory breaks down in the multi-TeV WIMP mass regime, our
resummed results exhibit a convergent  perturbative expansion.   The
leading effect relative to tree level is represented by a universal
Sudakov suppression, which at $M = 3 \, {\rm TeV}$ implies a resummed
cross section that is reduced by a factor $ \sim 3$.

In a forthcoming paper~\cite{future}, we will examine observational consequences in more detail, including the computation for triplet fermion annihilation and reinterpretation of constraints on this process using theoretically reliable cross sections.   This work demonstrates that accounting for large logarithms through resummation is necessary for robust predictions of the heavy WIMP annihilation cross section -- this is of clear importance in order to compare theory and indirect detection experiments.

\section*{Note Added}
While this work was in the final stages of preparation, \cite{Baumgart:2014vma} appeared which provides some partial results on resummation neglecting the effects of electroweak symmetry breaking for heavy WIMP dark matter.  We also became aware of another work on a similar topic \cite{Ovanesyan:2014fwa}, which is to appear soon.

\section*{Acknowledgements} We thank Thomas Becher, Matthew Dolan,
Matthias Neubert, Michael Peskin, Jon Walsh, Hua-Xing Zhu for insightful discussions.  
In particular, we are grateful to Mariangela
Lisanti, Aaron Pierce, and Tracy Slatyer for permission to use the
Sommerfeld enhancement code developed for  \cite{Cohen:2013ama}.  MB
is supported by the Alexander von Humboldt Foundation.  MB thanks
SLAC, where some of this research was performed, for support and
hospitality. TC is supported  by DoE contract number DE-AC02-76SF00515
and by an LHC Theory Initiative Postdoctoral Fellowship, under the
National Science Foundation grant PHY-0969510.   TC thanks the KITP in
Santa Barbara where some of this research was performed and for the
support from the National Science Foundation under Grant No. NSF
PHY11-25915.   TC also thanks the MITP in Mainz where additional work
was performed.  RH is supported by the DOE Grant No. DE-FG02-13ER41958.  
TC and RH thank the Aspen Center for Physics where this  work was initiated for
the support under NSF Grant No.~NSF PHY10-66293.  MS acknowledges
support from a Bloomenthal Fellowship at the University of Chicago and
from the  Office of Science, Office of High Energy Physics, of the
U.S.\ Department of Energy under contract DE-AC02-05CH11231.

\end{fmffile}
\end{spacing}

\pagebreak

\begin{spacing}{1.1}
\bibliography{SCET4WIMPs}
\bibliographystyle{utphys}
\end{spacing}

\end{document}